\documentclass[11pt,a4paper]{article}

\usepackage{epsfig}
\usepackage{amsmath,amsfonts,amssymb}
\usepackage{t1enc}
\usepackage{cite}

\newcommand{\kfac}{k}
\newcommand{\TTWH}{T \bar T \, (WH)}
\newcommand{\TTHH}{T \bar T \, (HH)}
\newcommand{\TTZH}{T \bar T \, (ZH)}
\newcommand{\TTWZ}{T \bar T \, (WZ)}
\newcommand{\TTZZ}{T \bar T \, (ZZ)}
\newcommand{\TTWZnoH}{T \bar T \, (WZ;H\!\!\!\!\!\!\! \not \,\,\,\,\,)}
\newcommand{\TTZZnoH}{T \bar T \, (ZZ;H\!\!\!\!\!\!\! \not \,\,\,\,\,)}
\newcommand{\ttnj}{t \bar t nj}
\newcommand{\ttHnj}{t \bar t H nj}
\newcommand{\ptlep}{p_t^\mathrm{lep}}
\newcommand{\ptmax}{p_t^{j,\mathrm{max}}}
\newcommand{\ptbmax}{p_t^{b,\mathrm{max}}}
\newcommand{\ptbmaxx}{p_t^{b,\mathrm{max2}}}
\newcommand{\ptlmax}{p_t^{j,\mathrm{max}}}
\newcommand{\ptlmaxx}{p_t^{j,\mathrm{max2}}}

\newcommand{\ptmiss}{p_t\!\!\!\!\!\!\!\! \not \,\,\,\,\,\,}
\newcommand{\mThad}{m_T^\mathrm{had}}
\newcommand{\mTlep}{m_T^\mathrm{lep}}
\newcommand{\mthad}{m_t^\mathrm{had}}
\newcommand{\mtlep}{m_t^\mathrm{lep}}
\newcommand{\mtrec}{m_t^\mathrm{rec}}
\newcommand{\mwhad}{M_W^\mathrm{had}}
\newcommand{\mhrec}{M_H^\mathrm{rec}}
\newcommand{\noH}{H\!\!\!\!\!\!\! \not \,\,\,\,\,}
\providecommand{\openone}{\leavevmode\hbox{\small1\kern-3.8pt\normalsize1}}

\begin{document}

\begin{center}
\begin{Large}
{\bf Light Higgs boson discovery from fermion mixing}
\end{Large}

\vspace{0.5cm}
J. A. Aguilar--Saavedra \\[0.2cm] 
{\it Departamento de Física Te\'orica y del Cosmos and CAFPE, \\
Universidad de Granada, E-18071 Granada, Spain} \end{center}

\begin{abstract}
We evaluate the LHC discovery potential for a light Higgs boson in $t \bar t H
\, (\to \ell \nu b \bar b b \bar b jj)$ production, within the
Standard Model and if a new $Q=2/3$ quark singlet $T$ with a moderate
mass exists. In the latter case, $T$ pair production with decays 
$T \bar T \to W^+ b \, H \bar t / H t \, W^- \bar b \to W^+ b W^- \bar b H$
provides an important additional source of Higgs bosons giving the same
experimental signature, and other decay modes
$T \bar T \to H t \, H \bar t \to W^+ b \, W^- \bar b$ $H H$,
$T \bar T \to Z t \, H \bar t / H t \, Z \bar t \to W^+ b W^- \bar b H Z$
further enhance this signal.
Both analyses are carried out with particle-level
simulations of signals and backgrounds, including
$t \bar t$ plus $n=0,\dots,5$ jets which constitute the main background by
far.
Our estimate for SM Higgs discovery in $t \bar t H$ production,
$0.4 \sigma$ significance for $M_H = 115$ GeV and an integrated luminosity
of 30 fb$^{-1}$, is similar to the most recent ones by CMS which also include
the full $\ttnj$ background.
We show that, if a quark singlet with a mass $m_T = 500$ GeV exists, the
luminosity required for Higgs discovery in this final state is
reduced by more than two orders of magnitude, and $5\sigma$ significance can be
achieved already with 8 fb$^{-1}$. 
This new Higgs signal will not be seen unless we look for it: with
this aim, a new specific final state reconstruction method is presented.
Finally, we consider the sensitivity to search for $Q=2/3$ singlets.
The combination of these three decay modes allows to discover a 500 GeV quark
with 7 fb$^{-1}$ of luminosity.
\end{abstract}

\section{Introduction}

The discovery of the Higgs boson is one of the main goals of the Large Hadron
Collider (LHC). Our present understanding of electroweak symmetry breaking in
the Standard Model (SM) relies on the existence of at least one of such scalar
particles \cite{higgs}, whose mass is however not predicted.
Direct searches at LEP have placed the limit $M_H > 114.4$ GeV on the mass of a
SM-like Higgs, with a 95\% confidence level (CL) \cite{hlimit}. Actually, data
taken from the ALEPH collaboration showed an excess of events over the SM
background consistent with a 115 GeV Higgs boson, but these results were not
confirmed by the other LEP collaborations. There is some theoretical
prejudice leading us to believe in the existence of a Higgs boson not much
heavier than this direct bound. Precision electroweak data seem to indicate its
existence, with a best-fit value of $M_H =
91^{+45}_{-32}$ GeV for its mass \cite{hfit} if the SM is assumed. On the other
hand, the Higgs boson must be lighter than around 1 TeV
if the SM is required to remain perturbative up to the unification scale
\cite{hlimit2}.

There is a vast Higgs search program at LHC, including various
production processes and the decay channels relevant in each mass range
\cite{tdr,CMS}. Most analyses focus on the search of a SM-like Higgs boson.
For masses $M_H \lesssim 130$ GeV the decay $H \to b \bar b$
dominates, with a branching fraction around $0.7$. However, the most important
production process $g g \to H$ is not visible in this channel due to the
enormous
QCD background. One has then to fall back either on rare decay modes,
production processes in association with extra particles, or both. 
One example is the production together with a $t \bar t$ pair, with
$H \to b \bar b$ and semileptonic decay of $t$, $\bar t$.
Further examples are
$g g \to H$ followed by $H \to \gamma \gamma$ (which has a branching
ratio around 0.2\%), or associate production $t \bar t H$, $W H$, $Z H$ with
$H \to \gamma \gamma$.
Simulations performed by the ATLAS collaboration \cite{ttHupd1,ttHupd2}
estimated that $t \bar t H$ with $H \to b \bar b$ allows to 
reach $5 \sigma$ significance for a 120 GeV Higgs boson with an integrated
luminosity of 100 fb$^{-1}$, while very recent results from CMS
\cite{cmshiggs}, with a more realistic background calculation, considerably
lower
these expectations: even in the ideal case of no systematic uncertainties, $5
\sigma$ significance could only be possible with $\sim 180$ fb$^{-1}$ (combining
several decay channels of the $t \bar t$ pair).
Hence, discovery of $t \bar tH$, with $H \to b \bar b$, seems unfeasible.
However, the combination of
$H$, $t \bar t H$, $W H$ and $Z H$ production, with $H \to \gamma \gamma$,
is expected to give $5 \sigma$ already with 60 fb$^{-1}$, providing also
a relatively precise measurement of the Higgs mass. Vector boson fusion
(VBF) processes $q q \to q' q' H$, with $H \to W^+ W^- \to \ell^+ \nu \ell^{'-}
\nu$, provide a similar sensitivity \cite{VBF}.

For larger Higgs masses the prospects are better. For $130 \lesssim M_H \lesssim
2 M_W$, $g g \to H$ production with decay $H \to Z Z^* \to
\ell^+ \ell^- \ell^{'+} \ell^{'-}$ provides a very clean experimental signature
of four charged leptons. A Higgs particle with $M_H = 130$ GeV may be
detected in this channel with 15 fb$^{-1}$, and for $M_H = 150$ GeV the
luminosity required is reduced to 3 fb$^{-1}$. VBF processes are also
interesting in this mass range, allowing to discover the Higgs with 12.9
fb$^{-1}$ for $M_H = 130$ GeV and 3.5 fb$^{-1}$ for $M_H = 150$ GeV \cite{VBF}.
For slightly larger masses, $2 M_W \lesssim M_H < 2 M_Z$,
the $H \to ZZ$ mode gets very suppressed due to the appearance of the on-shell
decay $H \to W^+ W^-$. Two signals are interesting in this range:
$g g \to H$, with $W^+ W^- \to \ell^+ \nu \ell^{'-} \nu$,
giving $5 \sigma$ significance for a luminosity around 4 fb$^{-1}$ \cite{WW160},
and again VBF processes, with leptonic or semileptonic decays of the $W$ pair,
which improve this result giving the same sensitivity for 2 fb$^{-1}$.
For masses larger than $2 M_Z$, the mode $H \to Z Z$ is possible with
both $Z$ bosons on their mass shell. This channel alone can signal the
existence of a Higgs boson with a luminosity ranging from 2.6 fb$^{-1}$ for $M_H
= 200$ GeV to 32 fb$^{-1}$ for $M_H = 600$ GeV \cite{ZZupd}. Larger masses up
to approximately 1 TeV can be probed combining different channels.

In SM extensions these production mechanisms can be enhanced or suppressed,
and new ones may appear. In this work we analyse in detail a new production
mechanism \cite{paco}, possible when the top quark mixes with a new $Q=2/3$
singlet. Such particles appear in Little Higgs models
\cite{lhiggs}, extra-dimensions \cite{extrad}, and grand unified
theories \cite{frampton}. They can be produced in pairs at LHC, through standard
QCD interactions, with a large cross section for moderate masses of few hundreds
of GeV. Their decays are determined by their mixing with SM quarks, which (by
theoretical considerations and experimental constraints) is expected to be
largest with the third generation. In particular, their decays to $Ht$ occur
with a branching ratio close to 25\% for $M_H \ll m_T$. This possibility
would be especially welcome, since it increases the observability of a Higgs
boson in the mass region $M_H \lesssim 130$ GeV where its detection is more
difficult. For definiteness, we will assume $M_H = 115$ GeV, though the results
are rather insensitive to the Higgs mass, as long as
the main decay channel is $H \to b \bar b$. The
largest cross section corresponds to
\begin{equation}
g g, q \bar q \to T \bar T \to W^+ b \, H \bar t / H t \, W^- \bar b 
\to W^+ b \, W^- \bar b \, H \,,
\label{ec:signal}
\end{equation}
with semileptonic decay of the $W$ pair and $H \to b \bar b$.
It gives the same experimental signature
$\ell \nu b \bar b b \bar b jj$ as SM $t \bar t H$ production but the
kinematics is rather different. Two further processes
contribute to the Higgs signal,
\begin{align}
& g g, q \bar q \to T \bar T \to H t \, H \bar t  
\to W^+ b \, W^- \bar b \, H H \,, \nonumber \\
& g g, q \bar q \to T \bar T \to Z t \, H \bar t / H t \, Z \bar t 
\to W^+ b \, W^- \bar b \, H Z \,,
\label{ec:signal2}
\end{align}
yielding the same final state, or the same state plus two jets,
when the extra Higgs and $Z$ boson decay $H \to b \bar b,c \bar c$,
$Z \to q \bar q,\nu \bar \nu$. In this work we
compare the discovery potential in this final
state within the SM (in which case the only signal is $t \bar t H$) and with a
new singlet $T$, assuming for its mass a reference value $m_T = 500$ GeV. 
It has been shown that such a particle could be seen at LHC in a short time,
through its decays $T \bar T \to W^+ b W^- \bar b$ \cite{plb}.\footnote{
$Q=2/3$ singlets with masses up to 1.1 TeV can be discovered at LHC in this
channel, for three years at the high luminosity run (100 fb$^{-1}$ per year).
For $Q=-1/3$ singlets the discovery reach is very similar \cite{unel}.} 
Experimental
search in the $\ell \nu b \bar b b \bar b jj$ final state would improve the
statistical
significance of the $T$ signal and, what is perhaps even more important,
it would allow a prompt discovery of the Higgs boson. 

We remark that, in contrast to what happens with a fourth sequential generation
\cite{4gen}, a quark singlet contributes very little to $g g \to H$ in general,
due to its tiny Yukawa coupling obtained by mixing with the top quark. The
amplitudes for $g g \to H$ mediated by $T$ and top quarks, relative to the SM
one (involving the top quark only), can be written as
\begin{eqnarray}
\frac{A(T)}{A(t)_\text{SM}} & = & \frac{y_{HTT}}{y_{Htt}|_\text{SM}}
  \left[ \frac{I(m_T^2/M_H^2)}{I(m_t^2/M_H^2)} \right]
= \frac{m_T X_{TT}}{m_t}
  \left[ \frac{I(m_T^2/M_H^2)}{I(m_t^2/M_H^2)} \right] \,, \notag \\
\frac{A(t)}{A(t)_\text{SM}} & = & \frac{y_{Htt}}{y_{Htt}|_\text{SM}}
= X_{tt}
\end{eqnarray}
being $X_{TT}$, $X_{tt}$ mixing factors (see next section for details) and $I$
a loop function.
The ratio in brackets is very close to unity for a light Higgs, and takes the
value $0.977$ for $M_H = 115$ GeV, $m_T = 500$ GeV. With typical
values $X_{TT} \simeq 0.04$, $X_{tt} \simeq 0.96$ for the mixing factors,
for $m_T = 500$ GeV the $T$ amplitude is about 9 times smaller than the SM
one, and the top quark contribution is reduced by a factor $0.96$.
We also note that in particular SM extensions including $Q=2/3$ singlets
other processes and/or channels may be enhanced or suppressed.
An interesting example takes place in Little Higgs models, where the
$g g \to H$ cross section may be suppressed but the branching ratio for
$H \to \gamma \gamma$ can increase in some regions of parameter space, due to
the extra contribution of the new fermions to the effective
$H \gamma \gamma$ vertex \cite{cpyuan}. We finally note that in models with
one (or more) $Q=-1/3$ singlet $B$ there are large Higgs signals from $B \bar B$
production and decay $B \to H b$, giving different final states from
the ones studied here \cite{paco}.

\section{Summary of the model}
\label{sec:2}

\begin{figure}[h*]
\begin{center}
\epsfig{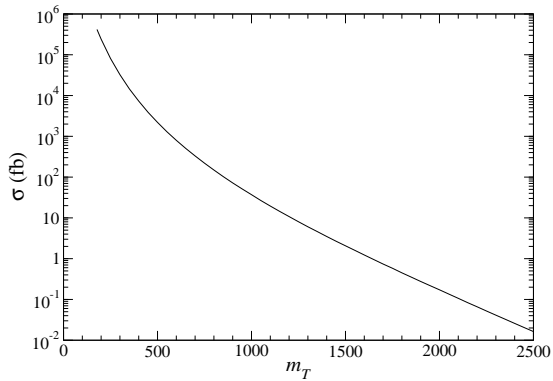}
\caption{Total production cross section for $gg,q \bar q \to T \bar T$
for different $T$ masses.}
\label{fig:mass-cross}
\end{center}
\end{figure}

SM extensions with vector-like quarks under $\text{SU}(2)_L$ have been
introduced before, and their phenomenology has been extensively explored
\cite{paconpb,london,barger,largo}. Here we will briefly recall the main
features of a SM extension with a $Q=2/3$ quark singlet, summarising the most
relevant points for this work.
The addition of two $\mathrm{SU}(2)_L$ singlet fields $T^0_{L,R}$ to the quark
spectrum modifies the weak and scalar interactions involving $Q=2/3$ quarks, but
does not affect strong and electromagnetic interactions.
(We denote weak eigenstates with a zero superscript, to distinguish them from 
mass eigenstates which do not bear superscripts.)
Thus, the new $Q=2/3$ mass eigenstate $T$ can be produced in pairs in $pp$
collisions via QCD interactions like the top quark. The production cross
section, plotted in Fig.~\ref{fig:mass-cross}, decreases with $m_T$ but is
sizeable for $T$ masses of several hundreds of GeV. For our evaluations we will
take $m_T = 500$ GeV, well above the present limit from Tevatron
$m_T \geq 258$ GeV at 95\% CL \cite{cdf}.\footnote{This limit assumes
$\mathrm{Br}(T \to W^+ b) = 1$. The new eigenstate can also decay $T \to Z t$,
$T \to H t$ (see below), but these two channels are kinematically forbidden for
$m_T = 258$ GeV.}

\begin{figure}[h*]
\begin{center}
\epsfig{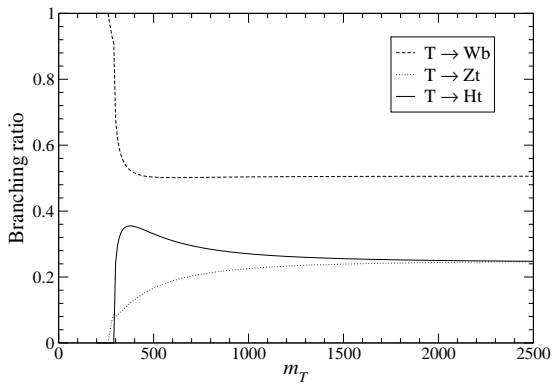} \\
\caption{Branching ratios for $T \to W^+ b$, $T \to Zt$, $T \to Ht$, for
different $T$ masses.}
\label{fig:mass-BR}
\end{center}
\end{figure}

The decay of the new quark takes place through electroweak and scalar
interactions. Using standard notation, these interactions read
\begin{eqnarray}
\mathcal{L}_W & = & - \frac{g}{\sqrt 2} \left[ \bar u \gamma^\mu V P_L  d
  \; W_\mu^+ + \bar d \gamma^\mu V^\dagger P_L u \; W_\mu^- \right] \,,
  \nonumber \\
\mathcal{L}_Z & = & - \frac{g}{2 c_W} \bar u \gamma^\mu \left[ X P_L
  - \frac{4}{3} s_W^2 \openone_{4 \times 4}  \right] u \; Z_\mu \,, \nonumber \\
\mathcal{L}_H & = & \frac{g}{2 M_W} \,
\bar u \left[ \mathcal{M}^u X P_L  + X \mathcal{M}^u P_R \right] u
 \;  H \,,
\label{ec:1}
\end{eqnarray}
where $u=(u,c,t,T)$, $d=(d,s,b)$ and $P_{R,L} = (1 \pm \gamma_5)/2$.
The extended Cabibbo-Kobayashi-Maskawa (CKM) matrix $V$ is of dimension
$4 \times 3$, $X = V V^\dagger$ is a non-diagonal $4 \times 4$ matrix and
$\mathcal{M}^u$ is the $4 \times 4$ diagonal up-type quark mass matrix.
The new mass eigenstate $T$ is expected to couple mostly with third
generation quarks $t$, $b$, because $T_L^0$, $T_R^0$ preferably mix with
$t_L^0$, $t_R^0$, respectively, due to the large top quark mass. $V_{Tb}$ is
mainly constrained by the contribution of the new quark to the $\mathrm{T}$
parameter \cite{largo}. For $m_T = 500$ GeV, the most recent
value $\mathrm{T} = -0.03 \pm 0.09$ \cite{pdb} implies 
$|V_{Tb}| \leq 0.17$ with a 95\% CL.
Mixing of $T_L^0$ with $u_L^0$, $c_L^0$, especially with the latter, is very
constrained by parity violation experiments and the measurement of $R_c$ and
$A_\mathrm{FB}^{0,c}$ at LEP, respectively \cite{london,prl},
implying small $X_{uT}$, $X_{cT}$.
The charged current couplings with $d,s$ must be small as well,
$|V_{Td}|, |V_{ts}| \sim 0.05$, because otherwise the new quark would
give large loop contributions to kaon and $B$ physics observables \cite{largo}.
Therefore, $|V_{Td}|,|V_{Ts}| \ll |V_{Tb}|$ and
$|X_{uT}|, |X_{cT}| \ll |X_{tT}|$. The couplings of the $t,T$ quarks can be
expressed in terms of the charged current coupling $V_{Tb}$,
\begin{eqnarray}
|V_{tb}|^2 & \simeq & 1-|V_{Tb}|^2 \,, \notag \\
X_{tt} & \simeq & 1-|V_{Tb}|^2 \,, \notag \\
X_{TT} & \simeq & |V_{Tb}|^2 \,, \notag \\
|X_{tT}|^2 & \simeq & |V_{Tb}|^2 (1-|V_{Tb}|^2) \,.
\label{ec:coups}
\end{eqnarray}
As it has been mentioned above, the $T \bar T$ cross section is
independent of $V_{Tb}$ and, as we will see below, branching ratios are
independent too. The only place where this mixing appears is the total $T$
width, which is much smaller than the experimental resolution for the $T$ mass.
Thus, $V_{Tb}$ has no influence at all in our results. For definiteness,
we have taken for our evaluations a coupling $V_{Tb} = 0.2$. This value
is slightly above the most recent 95\% limit from the $\mathrm{T}$ parameter
(and compatible with the previous one, $|V_{Tb}| \leq 0.26$). For this coupling,
the Yukawa coupling of the top quark is $y_{Htt} = (m_t/2 M_W) \, X_{tt}$,
reduced by a factor 0.96 with respect to its SM value, and the Yukawa of the
new quark is very small, $y_{HTT} = (m_T/2 M_W) \, X_{TT}$ with $X_{TT} = 0.04$.
The relevant decays of the new
quark are $T \to W^+ b ,\, Zt ,\, Ht$, with partial widths
\begin{align}
\Gamma(T \to W^+ b) & = \frac{\alpha}{16 \, s_W^2} |V_{Tb}|^2
  \frac{m_T^3}{M_W^2} \left[ 1-3 \frac{M_W^4}{m_T^4} + 2 \frac{M_W^6}{m_T^6}
  \right] \,, \nonumber \\ 
\Gamma(T \to Z t) & = \frac{\alpha}{32 s_W^2 c_W^2} |X_{tT}|^2
  \frac{m_T}{M_Z^2} \lambda(m_T,m_t,M_Z)^{1/2} \nonumber \\
  & \times  \left[ 1 + \frac{M_Z^2}{m_T^2}
  - 2  \frac{m_t^2}{m_T^2} - 2  \frac{M_Z^4}{m_T^4}  + \frac{m_t^4}{m_T^4}
  + \frac{M_Z^2 m_t^2}{m_T^4} \right] \,, \nonumber \\
\Gamma(T \to H t) & = \frac{\alpha}{32 s_W^2} |X_{tT}|^2
 \frac{m_T}{M_W^2} \lambda(m_T,m_t,M_H)^{1/2} \nonumber \\
  & \times  \left[ 1 + 6 \frac{m_t^2}{m_T^2} - \frac{M_H^2}{m_T^2} 
  + \frac{m_t^4}{m_T^4} - \frac{m_t^2 M_H^2}{m_T^4} \right] \,,
\label{ec:3}
\end{align}
with
\begin{equation}
\lambda(m_T,m_t,M) \equiv (m_T^4 + m_t^4 + M^4 - 2 m_T^2 m_t^2 
- 2 m_T^2 M^2 - 2 m_t^2 M^2)
\end{equation}
a kinematical function. The two couplings $V_{Tb}$,
$X_{tT}$ involved in the decays are approximately equal (see
Eq.~(\ref{ec:coups})). Since the three partial widths are proportional to
$|V_{Tb}|^2$, the branching ratios only depend on $m_T$ and $M_H$. They are
plotted in Fig.~\ref{fig:mass-BR} for a fixed value $M_H = 115$ GeV.
For $m_T = 500$ GeV, we have $\mathrm{Br}(T \to W^+ b) = 0.503$,
$\mathrm{Br}(T \to Z t) = 0.166$, $\mathrm{Br}(T \to H t) = 0.331$.
(The total $T$ width is $\Gamma_T = 3.115$ for $V_{Tb} = 0.2$.)
Decays $T \to Zt \to \ell^+ \ell^- W^+ b$, $\ell=e,\mu$ give 
a cleaner final state than $T \to W^+ b$, but with a branching ratio 10 times
smaller. The channel $T \to W^+ b$ ($\bar T \to W^- \bar b$)
gives the best discovery potential for the new quark in single $T$
\cite{azuelos} as well as in $T \bar T$ production \cite{plb}. The remaining
decay $T \to H t$ constitutes a copious source of Higgs bosons for moderate $T$
masses, for which the $T \bar T$ production cross section is large.
We point out that in the minimal SM extension where only one $Q=2/3$ singlet is
introduced these branching ratios are independent of the mixing, and the new
quark (provided it is not decoupled) always decays $T \to H t$ if $m_T >
m_t + M_H$. In models with extra interactions, decays to $W'$, $Z'$
bosons may occur, if kinematically allowed. If additional
scalars exist, mixing with the lightest one $H$ might also be suppressed, if
this Higgs is not SM-like.

\section{Signal and background simulation}
\label{sec:3}

Many SM and some new physics processes give or mimic the experimental signature
studied of a charged lepton, at least four $b$-tagged jets and two non-tagged
jets, plus missing energy. The relevant processes are calculated with
matrix-element-based Monte Carlo generators and fed into {\tt PYTHIA}
\cite{pythia} to include initial and final state radiation (ISR, FSR)
and pile-up, and perform hadronisation.
The main background is constituted by $t \bar t + n$ jet production.
It is calculated, with $n = 0,\dots,5$, with {\tt ALPGEN} \cite{alpgen}, using
the MLM prescription \cite{mlm} to avoid double counting of jet
radiation performed by {\tt PYTHIA}. {\tt ALPGEN} is also used to calculate
the production of $W$ and $Z$ bosons plus six jets, or a $b \bar b \,/\,
c \bar c$ pair and four jets.
New Monte Carlo
generators are developed for $T \bar T$,
$t \bar t H$, $t \bar t b \bar b$ and $t \bar t c \bar c$ (through QCD and
electroweak (EW) interactions) and $W b \bar b b \bar b$ production, plus other
processes obtained replacing the top quarks by heavy $T$ quarks. These
generators use the full resonant
tree-level matrix elements for the production and decay processes, namely
\begin{align}
g g, q \bar q & \to T \bar T \to W^+ b \, H \bar t / H t \, W^- \bar b 
 \to W^+ b W^- \bar b H \to f_1 \bar f_1' b \bar f_2 f_2' \bar b
 \;  b \bar b / c \bar c \,, \notag \\
g g, q \bar q & \to T \bar T \to H t \, H \bar t 
 \to W^+ b W^- \bar b H H \to f_1 \bar f_1' b \bar f_2 f_2' \bar b
 \;  b \bar b / c \bar c \;  b \bar b / c \bar c \,, \notag \\
g g, q \bar q & \to T \bar T \to Z t \, H \bar t / H t \, Z \bar t 
 \to W^+ b W^- \bar b H Z \to f_1 \bar f_1' b \bar f_2 f_2' \bar b
 \;  b \bar b / c \bar c \; q \bar q / \nu \bar \nu \,, \notag \\
g g, q \bar q & \to T \bar T \to W^+ b \, Z \bar t / Z t \, W^- \bar b 
 \to W^+ b W^- \bar b Z \to f_1 \bar f_1' b \bar f_2 f_2' \bar b \;  b \bar b
  / c \bar c \,, \notag \\
g g, q \bar q & \to T \bar T \to Z t \, Z \bar t
 \to W^+ b W^- \bar b Z Z \to f_1 \bar f_1' b \bar f_2 f_2' \bar b
 \;  q \bar q / \nu \bar \nu \; q \bar q / \nu \bar \nu \,, \notag \\
g g, q \bar q & \to t \bar t H  \to W^+ b W^- \bar b H \to f_1 \bar f_1' b
 \bar f_2 f_2' \bar b \;  b \bar b / c \bar c \,, \notag \\
gg,q \bar q & \to t \bar t b \bar b \to W^+ b W^- \bar b b \bar b \to 
  f_1 \bar f_1' b \bar f_2 f_2' \bar b \, b \bar b \,, \notag \\
gg,q \bar q & \to t \bar t c \bar c \to W^+ b W^- \bar b c \bar c \to 
  f_1 \bar f_1' b \bar f_2 f_2' \bar b \, c \bar c \,, \notag \\
q \bar q' & \to W^\pm b \bar b b \bar b \to f_1 \bar f_1' b \bar b b \bar b \,.
\label{ec:proc}
\end{align}
Matrix elements are calculated with {\tt HELAS} \cite{helas}, partly using
{\tt MadGraph} \cite{madgraph}. All finite width and spin effects are thus
automatically taken into account.
The colour flow information necessary for {\tt PYTHIA} is
obtained following the same method as in {\tt AcerMC} \cite{acermc}, {\em i.e.}
we randomly select the colour flow among the possible ones on an
event-by-event basis, computing the probabilities of such a configuration from
the matrix element (taking into account the diagrams contributing to such
configuration). Integration
in phase space is done with {\tt VEGAS} \cite{vegas}, modified following
Ref.~\cite{acermc}.
These generators (except $W b \bar b b \bar b$) have been checked against
{\tt ALPGEN} using the same parameters, structure functions and
factorisation scales, obtaining very good agreement. For our evaluations we take
$m_t = 175$ GeV, $m_b = 4.8$ GeV, $m_c = 1.5$ GeV (neglected in $W$
decays), $\alpha(M_Z) = 1/128.878$, $s_W^2 (M_Z) = 0.23113$, $\alpha_s(M_Z) =
0.127$ and run the coupling constants up to the the scale of the heavy
($t$ or $T$) quark.
Structure functions CTEQ5L \cite{cteq} are used, with
$Q^2 = \hat s$ the square of the partonic centre of mass energy. (For
{\tt ALPGEN} processes we select $Q^2 = M_{t,W,Z}^2 + p_{t_{t,W,Z}}^2$.)
Several representative total cross sections obtained (without decay branching
ratios nor phase space cuts)  can be found in Table~\ref{tab:cstot}, for
comparison with other generators. The total cross sections for
$t \bar t c \bar c$, $\ttnj$
with $n \geq 1$ and $W/Z +$ jets are numerically unstable due to collinear
singularities and not shown. This is not a problem for event generation, since
suitable kinematical cuts at the generator level (discussed below) can be
applied to stabilise the cross sections.

\begin{table}[h*]
\begin{center}
\begin{tabular}{cc}
Process & $\sigma_\text{tot}$ \\
$t \bar t$ ({\tt ALPGEN}) & 489 pb \\
$T \bar T$ & 2.14 pb  \\
$t \bar t H $ & 508 fb  \\
$t \bar t b \bar b$ & 8.65 pb \\
$t \bar t b \bar b$ EW & 773 fb  \\
$W b \bar b b \bar b$ & 303.4 fb 
\end{tabular}
\caption{Total cross sections for several processes studied.}
\label{tab:cstot}
\end{center}
\end{table}

In our analysis
we consider semileptonic decays of the $W^+ W^-$ pairs, and leptonic decays in
the production of $W/Z \,+$ jets. The main contributions come from $\ell = e,
\mu$,
but decays to $\tau$ leptons are included as well. Phase space cuts are applied
at the generator level in some processes to reduce statistical fluctuations and
improve the unweighting efficiency. The cuts applied are
\begin{align}
\ttnj & & & |\eta^j| \leq 2.5 \;,~ p_t^j \geq 20~\text{GeV} \;,~
\Delta R^{jj} \geq 0.4 \notag \\
t \bar t b \bar b, t \bar t c \bar c, W b \bar b b \bar b 
& & & |\eta^{b,c}| \leq 2.5 \;,~ p_t^{b,c} \geq 15~\text{GeV} \notag \\
W b \bar b jjjj, W c \bar c jjjj, W jjjjjj 
& & & |\eta^{\ell,b,j}| \leq 2.5 \;,~ p_t^{\ell} \geq 6~\text{GeV}
 \;,~ p_t^{b,j} \geq 15~\text{GeV} \;, \notag \\
& & &  \Delta R^{jj,bb} \geq 0.4 \;,~ \Delta R^{\ell j, \ell b} \geq 0.4 \notag
 \\
Z b \bar b jjjj, Z c \bar c jjjj, W jjjjjj 
& & & |\eta^{b,j}| \leq 2.5 \;,~ p_t^{\ell,\text{max}} \geq 6~\text{GeV} 
\;,~  p_t^{b,j} \geq 15~\text{GeV} \;, \notag \\
& & & \Delta R^{jj,bb} \geq 0.4
\,,
\end{align}
where $\eta$ is the pseudorapidity, $p_t$ the transverse momentum and 
$\Delta R \equiv \sqrt{(\Delta \eta)^2 + (\Delta \phi)^2}$ the lego-plot
distance.
The cross sections after decay, including generator cuts, can be read in
Table~\ref{tab:cs} for $\ell = e,\mu$. The $T \bar T$ processes in
Eqs.~(\ref{ec:proc}) will from now on be denoted according to the decay mode
as $\TTWH$, $\TTHH$, $\TTZH$, $\TTWZ$ and $\TTZZ$. Sum over charge conjugate
decays is always understood.

\begin{table}[htb]
\begin{center}
\begin{small}
\begin{tabular}{cccccccc}
Process & $\sigma$ & $\varepsilon$ & \quad \quad & Process & $\sigma$ & 
  $\varepsilon$ \\
$\TTWH$                & 173.6 fb & 6.3\% & &
   $t \bar t b \bar b$    & 564.9 fb & 4.7\% \\
$\TTHH$                & 44.38 fb & 19.3\% & &
   $t \bar t c \bar c$    & 630.5 fb & 0.65\% \\
$\TTZH$                & 50.0 fb & 8.5\% & &
   $t \bar t b \bar b$ EW & 60.31 fb & 4.8\% \\
$\TTWZ$                & 29.03 fb & 4.5\% & &
   $t \bar t c \bar c$ EW & 17.12 fb & 0.72\% \\
$\TTZZ$                & 14.07 fb & 2.8\% & &
   $Wjjjjjj$              & 69.85 pb & $\sim 7.4 \times 10^{-6}$ \\
$T \bar T b \bar b$    & 1.054 fb & 5.3\% & &
   $W b \bar b jjjj$      & 2.825 pb & 0.12\% \\
$t \bar t H$           & 118.7 fb & 4.7\% & &
   $W c \bar c jjjj$      & 3.279 pb & $\sim 0.015$\% \\
$t \bar t$             & 143.2 pb & 0.034\% & &
   $W b \bar b b \bar b$  & 2.587 fb & $\sim 3.4$ \% \\
$t \bar tj$            & 142.7 pb & 0.055\% & &
   $Z jjjjjj$             & 10.48 pb & $\sim 3.9 \times 10^{-6}$ \\
$t \bar t2j$           & 95.9 pb  & 0.085\% & &
   $Z b \bar b jjjj$      & 722.5 fb & 0.090\% \\
$t \bar t3j$           & 54.0 pb  & 0.12\% & &
   $Z c \bar c jjjj$      & 738.5 fb & $\sim 0.013$\% \\
$t \bar t4j$           & 27.4 pb  & 0.15\% & &
    \\
$t \bar t5j$           & 12.8 pb  & 0.19\% 
\end{tabular}
\end{small}
\caption{Cross section at the generator level and efficiency $\varepsilon$
for signal and
background processes in the decay channels with $\ell = e,\mu$. The
corresponding cross sections for final states with tau leptons
are approximately one half, with efficiencies $20-30$ times smaller.}
\label{tab:cs}
\end{center}
\end{table}

The generated events are passed through {\tt PYTHIA 6.403} as external
processes to
include ISR, FSR, pile-up and perform hadronisation.\footnote{In order to avoid
double counting, in the {\tt PYTHIA} simulation of
the $W/Z \,+$ 6 jets processes we turn off $b \bar b$ and $c \bar
c$ pair radiation, which are independently generated. Similarly,
for $t \bar t c \bar c$ and $W/Z + b \bar b \,/\, c \bar c \,+$ 4 jets
we turn off $b \bar b$ pair radiation. The radiation of extra jets in $\ttnj$
processes is vetoed following the MLM prescription.}
We use the standard {\tt PYTHIA} settings
except for $b$ fragmentation, in which we use the Peterson parameterisation with
$\epsilon_b=-0.0035$ \cite{epsb}. For pile-up we take 4.6 events in average,
corresponding to a luminosity of $2 \times 10^{33} \;\text{cm}^{-2}
\;\text{s}^{-1}$. Tau leptons in the final state are decayed
using {\tt TAUOLA} \cite{tauola} and {\tt PHOTOS} \cite{photos}.
 A fast detector simulation {\tt ATLFAST 2.60} 
\cite{atlfast}, with standard settings, is used for the modelling of the ATLAS
detector. We reconstruct jets using a cone algorithm with
$\Delta R = 0.4$. This cone size has proved to be the most adequate for top
physics studies \cite{int}, providing very good agreement between fast and full
simulations for reconstructed quantities \cite{marsella}. We do not apply
trigger inefficiencies and assume a perfect
charged lepton identification.
The package {\tt ATLFASTB} is used to recalibrate jet energies
and perform $b$ tagging, for which we select a 60\% efficiency at the low
luminosity run, with nominal rejection factors of 93 for light jets and 6.7
for charm, and $p_t$-dependent corrections. These efficiencies are in agreement
with those obtained from full simulations \cite{marsella2}, and comparable to
the ones expected at CMS \cite{CMS2}.

The hadronised events are required to fulfill these two criteria: (a) the
presence of one (and only one) isolated charged lepton, which must have
transverse momentum $p_t \geq 25$ GeV (for electrons), $p_t \geq 20$ GeV (for
muons) and  $|\eta| \leq 2.5$; (b)
at least six jets with $p_t \geq 20$ GeV, $|\eta| \leq 2.5$, with at least
four $b$ tags and two untagged jets.
The charged leptons provide a trigger for the events \cite{trigger}.
Signal and background efficiencies after these requirements are
shown in Table~\ref{tab:cs}. We notice the higher acceptance for
the $\TTHH$
process, with six $b$ quarks in the final state when both Higgs bosons decay to
$b \bar b$, and for $\TTZH$, where sometimes two $b$ quarks are produced in the
$Z$ decay. We also point out the growing efficiency of the $\ttnj$ processes
with increasing multiplicity.

Finally, we must note that our calculation of the $W b \bar b b \bar b jj$
background, with $W b \bar b b \bar b$ production at the generator level and
extra jet radiation performed by {\tt PYTHIA}, must be regarded as an estimate.
The reason is that in $W b \bar b b \bar b$ only $q \bar q'$ scattering
processes are
involved, while gluon fusion contributes to $W b \bar b b \bar b jj$.
At any rate, this background turns out to be completely negligible.
$Z b \bar b b \bar b$ production has an even smaller
cross section and we have not included it in our calculations. 
We have investigated $t \bar t b \bar b b \bar b$ production with {\tt ALPGEN},
which might be important if five or more $b$ tags are required.
The cross section (with the same cuts used before) is of 0.54 fb. Assuming a
similar detection efficiency as for $t \bar t b \bar b$, the
requirement of five tagged jets reduces the cross section to one
event for 30 fb$^{-1}$ (and 0.3 events with 6 $b$ tags). One may also
think about $T \bar T H$ production, with $T \bar T \to W^+ b W^- \bar b$, also
contributing to the final state studied. This process is irrelevant
due to the small Yukawa coupling of the $T$ quark.

\section{Higgs boson discovery}
\label{sec:4}

We simulate events for an integrated luminosity of 30 fb$^{-1}$, which can be
collected in three years at the low luminosity phase. For some background
processes the
number of events simulated corresponds to the cross section obtained from the
generator scaled by a $\kfac$ factor, to take into account higher order
contributions with extra jets. (This $\kfac$ factor accounting for higher
multiplicity processes must not be confused with a $K$ factor to take radiative
corrections into account.)
For the main background, $\ttnj$ production, higher
order processes are explicitly calculated, and $\kfac$ factors are not included
except for $N=5$, where we set $\kfac = 1.46$ to account for $t \bar t + 6$
jets. For $t \bar t b \bar b$ and $t \bar t c \bar c$, the $\kfac$ factor is 
estimated from the $\ttnj$ cross sections as
$\kfac =  [\sigma(t \bar t 2j) + \dots + \sigma(t \bar t 6j)] / \sigma(t \bar t
2j) = 2.05$. For $W/Z$ plus jets we use the approximate prescription in
Ref.~\cite{plb}, which gives $\kfac = 2-3$. For all signals we conservatively set
$\kfac=1$. The reason for this will be explained later.
The number of events generated for each process can be read in
Table~\ref{tab:ang}. In the sums, the first term corresponds to final states
with $\ell=e,\mu$ and the second one to $\ell=\tau$, but in the following all
lepton channels will be summed.
A subtlety in the analysis is that when the singlet $T$ is introduced the
$Htt$, $Wtb$ and $Ztt$ couplings of the top quark are modified. This affects
electroweak $t \bar t b \bar b$ and $t \bar t c \bar c$ production in a
non-trivial way, and different samples (taking into account the corrections to
the couplings) must be generated and simulated. $t \bar t H$ production is
modified as well, with the Yukawa coupling of the top quark reduced by a factor
$X_{tt} < 1$. In our case, we have assumed
a large mixing $V_{Tb} = 0.2$, for which $|V_{tb}| = 0.98$ and $X_{tt} = 0.96$.
These processes are indicated with a ``$(T)$'' in Table~\ref{tab:ang}, where we
can observe that the effect of mixing is negligible for $t \bar t b \bar b$
and $t \bar t c \bar c$. A second issue to keep in mind is that
when studying the new physics signals associated to the $T$ quark we must
distinguish the cases where the Higgs boson is present or not (if not, the
branching ratios for $T \to W^+ b$ and $T \to Zt$ are larger). The latter are
denoted with a ``$\noH$''.

\begin{table}[t*]
\begin{center}
\begin{footnotesize}
\begin{tabular}{ccccccc}
& $N_0$ & $N$ &   \quad & & $N_0$ & $N$ \\
$\TTWH$                       & $5200^* + 2600^*$   & $329.8+9.2$ & &
   $t \bar t b \bar b$           & $34700 + 17400$     & $1648 + 35$ \\
$\TTHH$                       & $1330^* + 665^*$   & $256.5+6.2$ & &
   $t \bar t c \bar c$           & $38800 + 19300$     & $253 + 8$ \\
$\TTZH$                       & $1500^* + 750^*$   & $127.4+3.1$ & &
   $t \bar t b \bar b$ EW        & $3710^* + 1850^*$   & $178.1 + 3.2$ \\
$\TTWZ$                       & $871^* + 436^*$    & $39.5 + 0.7$ & &
   $t \bar t b \bar b$ EW ($T$)  & $3420^* + 1710^*$   & $170.7 + 2.9$ \\
$\TTWZnoH$                    & $1950^* + 975^*$   & $87.7 + 2.3$ & &
   $t \bar t c \bar c$ EW        & $1050^* + 525^*$    & $7.6 + 0.0$ \\
$\TTZZ$                       & $422^* + 211^*$     & $11.9 + 0.2$ & &
   $t \bar t c \bar c$ EW ($T$)  & $980^* + 490^*$     & $6.3 + 0.1$ \\
$\TTZZnoH$                    & $939^* + 470^*$    & $27.3 + 0.4$ & &
   $Wjjjjjj$                     & $5270000 + 2640000$ & $39+0$ \\
$T \bar T b \bar b$           & $31.6^* + 15.4^*$   & $1.7 + 0.0$ & &
   $W b \bar b jjjj$             & $168000+83900$      & $208+4$ \\
$T \bar T b \bar b \; (\noH)$ & $70.8^* + 34.6^*$    & $4.1 + 0.1$ & &
   $W c \bar c jjjj$             & $195000+97400$      & $29+1$ \\
$t \bar t H$                  & $3560^* + 1780^*$   & $166.0 + 4.3$ & &
   $W b \bar b b \bar b$         & $118 +59$           & $4+1$ \\
$t \bar t H$ ($T$)            & $3280^* + 1640^*$   & $152.9 + 3.8$ & &
   $Zjjjjjj$                     & $1020000 + 510000$  & $4+0$ \\
$t \bar t$                    & $4368000 + 2184000$ & $1475 + 23$ & &
   $Z b \bar b jjjj$             & $53600 + 26800$     & $48+5$ \\
$t \bar t j$                  & $4282000 + 2141000$ & $2370 + 48$ & &
   $Z c \bar c jjjj$             & $54800 + 27400$     & $7+1$ \\
$t \bar t 2j$                 & $2878000 + 1439000$ & $2443 + 42$ & &
    \\
$t \bar t 3j$                 & $1620000 + 810000$  & $1900 + 48$ & &
    \\
$t \bar t 4j$                 & $822000 + 411000$   & $1195 + 45$ & &
    \\
$t \bar t 5j$                 & $562000 + 281000$   & $1067 + 19$
\end{tabular}
\end{footnotesize}
\caption{For each process: number of events simulated $N_0$ and number of events
passing the pre-selection criteria $N$. The first terms in the sums correspond
to $\ell=e,\mu$, and the second ones to $\ell=\tau$. For some contributions
(marked with an asterisk) we have simulated at least $10 N_0$ events and
rescaled the result to 30 fb$^{-1}$, so as to reduce statistical fluctuations.}
\label{tab:ang}
\end{center}
\end{table}

The discovery potential for the Higgs boson crucially depends on systematic
errors.
The uncertainty in the background normalisation makes it difficult to detect
the presence of a Higgs boson with a measurement of the total $\ell \nu bbbbjj$
cross section. Naively, from the data in Table~\ref{tab:ang} one could conclude
that the statistical significance of the $t \bar t H$ signal, before applying
any kinematical cut, is
$S/\sqrt B = 170.3/\sqrt{13158.9} = 1.48\sigma$. However,
this estimate does not include the systematic uncertainty in the SM background
total cross section ({\em i.e.} the background normalisation). A detailed
calculation of systematic uncertainties is beyond the scope of this work.
They generally arise from two sources: (i) the theoretical uncertainty in cross
sections, due to higher loop contributions and uncertainty in parton
distribution functions, among others; (ii) the systematic uncertainty related to
the experimental detection ($b$ tagging, jet energy scaling, etc.). The former
can go up to 30--50\% for $\ttnj$ with large $n$, but they
are reducible with more accurate theoretical calculations and/or background
measurements (understanding to what extent they are
reduced probably requires real data). For the latter we assume a ``reference''
value of 20\%, close to the value $\sim 26\%$ obtained in Ref.~\cite{cmshiggs}
with a detailed analysis for the CMS detector. We replace the
estimator $\mathcal{S}_0 \equiv S/\sqrt B$ by 
\begin{equation}
\mathcal{S}_{20} \equiv S/\sqrt{B+(0.2 B)^2} \,,
\end{equation}
where $S$ is the excess of events over the expected background.
Incorporating systematic uncertainties in the previous example, we obtain a
much smaller (but more realistic) significance $\mathcal{S}_{20} = 0.064
\sigma$. We note that adding statistical and systematic uncertainties in
quadrature is not the only way to incorporate systematics into the significance.
Other possibilities exist, which are perhaps more correct from the
statistical point of view, but we use this one for simplicity and in order to
compare better with other studies.

In the following
we perform two different analyses of signals and backgrounds.
The first one is a ``standard'' analysis aiming to discover 
the Higgs boson in $t \bar t H$ production, in which we reconstruct the final
state to distinguish this signal from the SM background. In case
that a new quark $T$ exists, additional signal events will improve
the Higgs discovery potential. The second analysis specifically looks for a
Higgs boson produced in heavy quark decays, optimising the reconstruction for
this signal.

\subsection{Analysis I: $\boldsymbol{t \bar t H}$ reconstruction}

The reconstruction of the $t \bar t H$ signal is not done sequentially, but
rather all possible pairings for light and $b$ jets are tried, selecting the
one which best resembles the kinematics of this process.
We reconstruct the $W$ boson decaying hadronically (called ``hadronic'' $W$
boson) from a pair of untagged jets
$j_1$ and $j_2$. For the leptonic $W$, the missing transverse
momentum is assigned
to the neutrino, and its longitudinal momentum and energy are found requiring
that the invariant mass of the charged lepton and neutrino is the $W$ mass,
$(p_\ell + p_\nu)^2 = M_W^2$. This equation gives two real solutions in
most cases. In case there is no real solution (the discriminant of the
quadratic equation is negative) we set it to zero to obtain a solution.
This procedure gives reconstructed mass distributions almost indistinguishable
from the ones obtained using the collinear approximation, {\em i.e.} setting
$p_\nu^z = p_\ell^z$.

\begin{figure}[p]
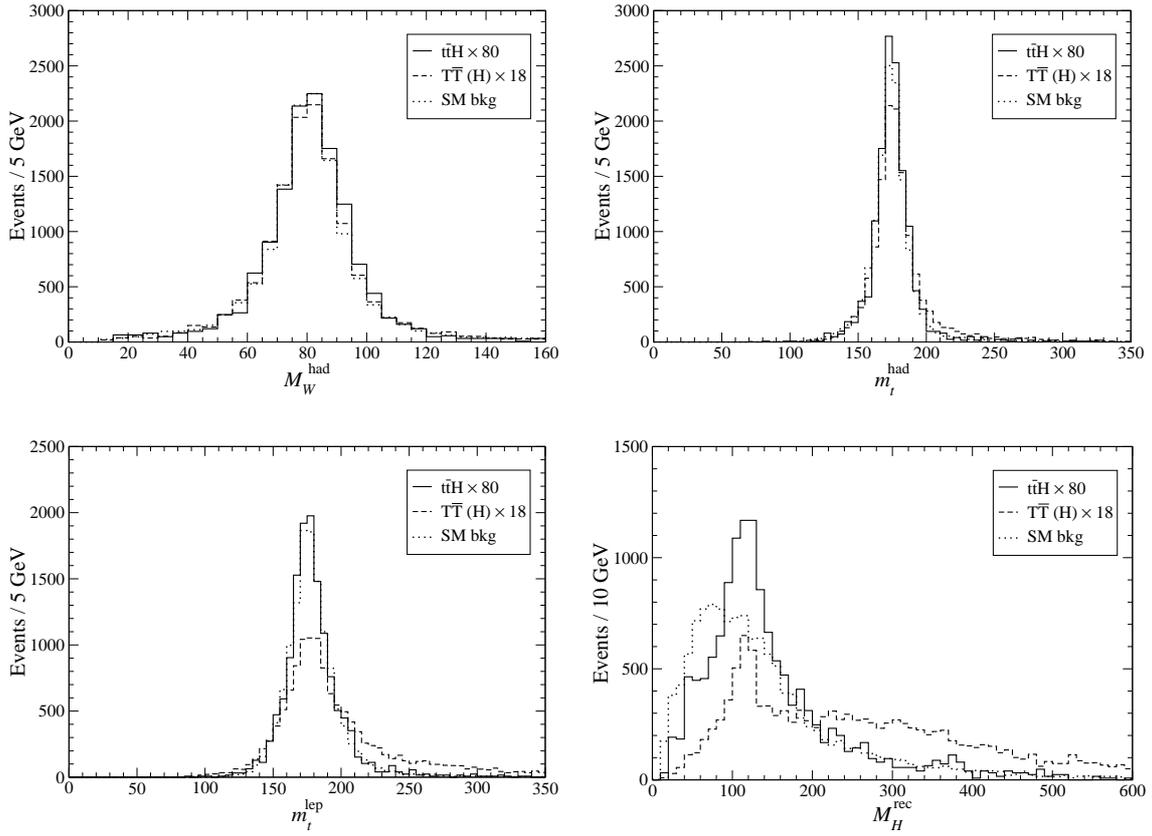

\begin{center}
\begin{tabular}{cc}
\epsfig{file=Figs/mWhad-anI.eps,height=5.2cm,clip=}
 & \epsfig{file=Figs/mthad-anI.eps,height=5.2cm,clip=} \\[0.4cm]
\epsfig{file=Figs/mtlep-anI.eps,height=5.2cm,clip=}
 & \epsfig{file=Figs/mH-anI.eps,height=5.2cm,clip=} \\[0.4cm]
\end{tabular}
\caption{Analysis I: Reconstructed masses of the hadronic $W$, the hadronic
and leptonic top quarks and the Higgs boson.}
\label{fig:anI}
\end{center}
\end{figure}

For each choice of $j_1$, $j_2$ and leptonic $W$ momentum, there are 12 possible
assignments of the four $b$ jets to the two $W$ bosons, to form the two top
quarks. (Around 9\% of the
signal events have five or more $b$ jets, in which case we select the four
with the highest transverse momentum.) Among all possibilities, we
select the one minimising the quantity
\begin{equation}
\Delta m^2 = 
\frac{(\mthad-m_t)^2}{S_t^2} + \frac{(\mtlep-m_t)^2}{S_t^2} +
\frac{(\mwhad-M_W)^2}{S_W^2} \,,
\end{equation}
where $\mthad$, $\mtlep$ and $\mwhad$ are the reconstructed masses of the
hadronic and leptonic top quarks, and the hadronic $W$, respectively. 
$S_t$ and $S_W$ are fixed parameters corresponding to the widths of the
reconstructed distributions, which are taken in this case to be equal,
$S_t = S_W = 10$ GeV. For the best combination, the two remaining (unpaired)
$b$ jets are assumed to originate from the Higgs boson decay,
whose momentum and invariant mass can then be reconstructed.
Kinematical cuts are not applied at this level.
The results are shown in Fig.~\ref{fig:anI}.
The reconstruction
works very well for the $t \bar t H$ signal, with sharp peaks for the
reconstructed masses $\mwhad$, $\mthad$ and $\mtlep$, and the Higgs mass
distribution mainly concentrated around the true value $M_H = 115$ GeV. 
Since the SM
background is dominated by $\ttnj$ production with two top
quarks, the invariant masses of the hadronic $W$ and the top pair are very well
reconstructed too. For the $T \bar T$ Higgs signal this reconstruction method is
not adequate, and the reconstructed Higgs mass spreads over a wider range.

The signal significance can be improved by simply performing a kinematical cut
on the Higgs reconstructed mass. Additionally, we perform a probabilistic
analysis (see appendix~\ref{ap:b}), involving the following
variables: 
\begin{itemize}
\item The light jet multiplicity $\mathcal{N}_\text{jet}$.
\item The smallest invariant mass of a $bb$ pair $m_{bb}^{(1)}$ \cite{ttHupd1},
among those involving the four jets with largest transverse momentum.
\item The sum of the transverse momenta of the two top quarks,
$p_t^\text{had} + p_t^\text{lep}$.
\item Angular quantities characterising the topology of the event: the azimuthal
angle and rapidity difference (i) between the two $b$ jets assigned
to the Higgs, $\Delta \phi_{bb}$ and $\Delta \eta_{b b}$;
(ii) between the Higgs and the closest (in $\Delta R$) top quark,
$\Delta \phi_{Ht}$ and $\Delta \eta_{H t}$; (iii)
between the two top quarks, $\Delta \phi_{tt}$ and $\Delta \eta_{tt}$.
\end{itemize}

\begin{figure}[p]
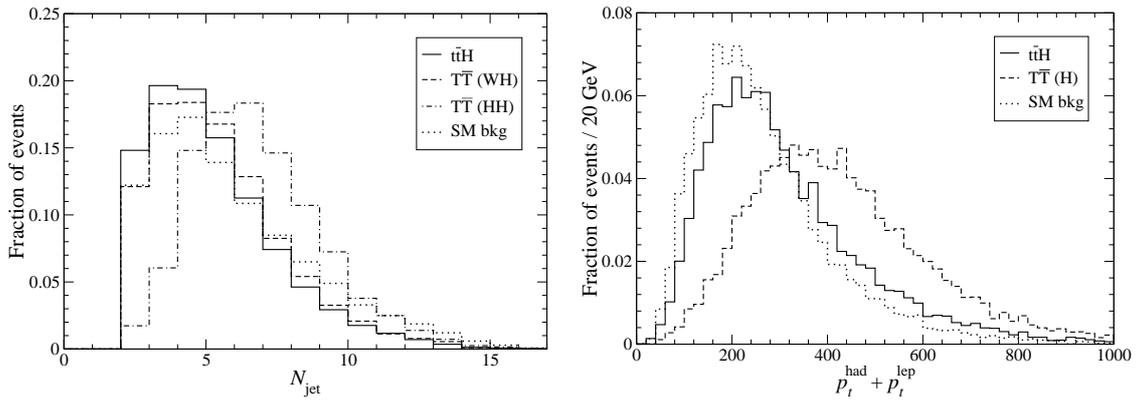

\begin{center}
\begin{tabular}{cc}
\epsfig{file=Figs/mult.eps,height=5.2cm,clip=}
 & \epsfig{file=Figs/pttsum_n-anI.eps,height=5.2cm,clip=} 
\end{tabular}
\caption{Analysis I: Normalised light jet multiplicity $\mathcal{N}_\text{jet}$
and variable
$p_t^\text{had} + p_t^\text{lep}$ (see the text), used in the probabilistic
analysis.  The jet multiplicities of the two main $T \bar T$ Higgs signals are
displayed separately for later convenience. The $p_t^\text{had} +
p_t^\text{lep}$ distribution for the $T \bar T$ Higgs signals is shown for
illustration, but not included in the probabilistic analysis as a separate event
class.}
\label{fig:anI-vars1}
\end{center}
\end{figure}

\begin{figure}[p*]
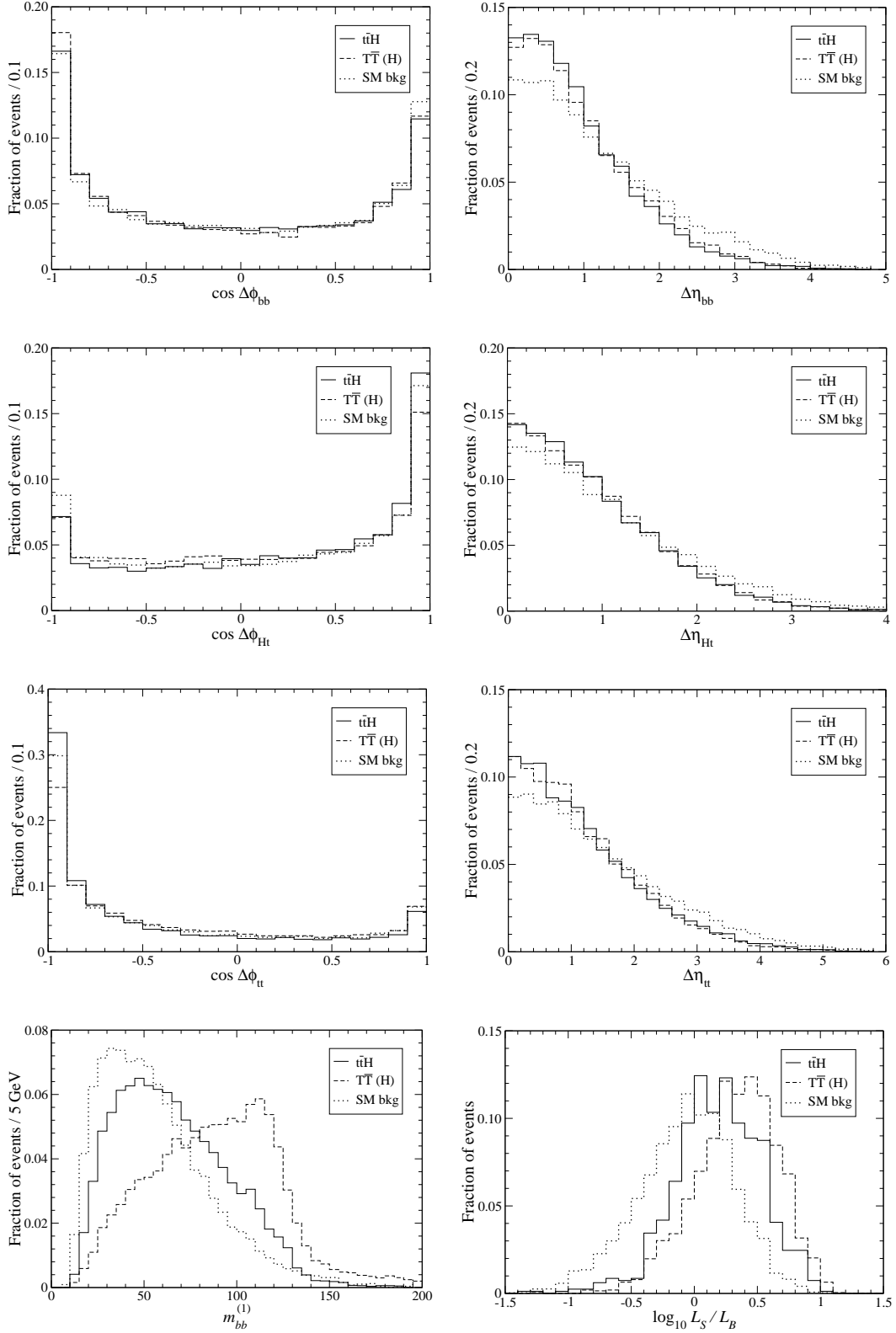

\begin{center}
\begin{tabular}{cc}
\epsfig{file=Figs/cPhbb_n-anI.eps,height=4.8cm,clip=}
 & \epsfig{file=Figs/dhbb_n-anI.eps,height=4.8cm,clip=} \\[0.4cm]
\epsfig{file=Figs/cPhHt_n-anI.eps,height=4.8cm,clip=}
 & \epsfig{file=Figs/dhHt_n-anI.eps,height=4.8cm,clip=} \\[0.4cm]
\epsfig{file=Figs/cPhtt_n-anI.eps,height=4.8cm,clip=}
 & \epsfig{file=Figs/dhtt_n-anI.eps,height=4.8cm,clip=}  \\[0.4cm]
 \epsfig{file=Figs/mbb1_n-anI.eps,height=4.8cm,clip=}
 &  \epsfig{file=Figs/logLSf_n-anI.eps,height=4.8cm,clip=}
\end{tabular}
\caption{Analysis I:  Normalised variables 
$\cos \Delta \phi_{bb}$, $\Delta \eta_{b b}$,
$\cos \Delta \phi_{Ht}$, $\Delta \eta_{H t}$,
$\cos \Delta \phi_{tt}$, $\Delta \eta_{tt}$ and $m_{bb}^{(1)}$
(defined in the text), used in the probabilistic analysis.
Log-likelihood function. 
The distributions for the $T \bar T$ Higgs signals are shown for illustration,
but not included in the probabilistic analysis as a separate event class.}
\label{fig:anI-vars2}
\end{center}
\end{figure}

These variables, plotted in
Figs.~\ref{fig:anI-vars1}, \ref{fig:anI-vars2} for the background and
reference signal samples (with more statistics), are not suitable for
kinematical cuts but help distinguish $t \bar t H$ production from the SM
background. Additional variables can be considered, but we have found no
improvement including them, and in some cases they reduce the discriminating
power of the likelihood functions (for a discussion see the appendix).
Using their distributions for $t \bar t H$ and the SM
background we build signal and background likelihood functions $L_S$, $L_B$.
The log-likelihood function $\log_{10} L_S/L_B$ is also
plotted in Fig.~\ref{fig:anI-vars2}. In the absence of systematic errors, the
highest statistical significance $\mathcal{S}_0 = S/\sqrt B$ would be achieved
with relatively loose cuts on $L_S/L_B$. But when one considers systematic
uncertainties, the highest
significance $\mathcal{S}_{20}$ is found for more strict cuts, which
reduce the background to few tens of events. For this purpose, we have found it
very useful to employ a hybrid event selection method,
in which we perform a simple cut
on the Higgs reconstructed mass and include the rest of the relevant variables
in the likelihood function. The kinematical cuts applied (not
fine-tuned but close to the optimal values) are
\begin{align}
& \log_{10} L_S/L_B \geq 0.75 \,, \notag \\
& 100~\text{GeV} \leq \mhrec \leq 140~\text{GeV} \,.
\label{ec:cut11}
\end{align}
The number of events corresponding to each process can be read in
Table~\ref{tab:anI}.
We point out that the inclusion of the light jet multiplicity as
a likelihood variable significantly reduces the $\ttnj$ background for larger
$n$. $W/Z$ plus jets is essentially eliminated for high $L_S$ values, even
without requiring explicitly a good $\mwhad$, $\mthad$ and $\mtlep$
reconstruction.
With these selection cuts
a statistical significance $\mathcal{S}_{20} = 0.39\sigma$ is found for 30
fb$^{-1}$.
This sensitivity is much lower than in previous ATLAS analyses
\cite{ttHupd1,ttHupd2} but similar to the most recent one by CMS,
$\mathcal{S}_{20} = 0.47\sigma$. \footnote{For a better comparison between both
results, this number has been obtained summing the number of events in the
electron and muon channel in Ref.~\cite{cmshiggs}, rescaling them
to 30 fb$^{-1}$ and assuming a 20\% background uncertainty}
However, it must be
noted that the CMS analysis uses full detector simulation, including the
electron and muon efficiencies not taken here into account. On the other hand,
the next-to-leading order cross section for $t \bar t H$ is used in that
analysis, which is 1.5 times larger than the one taken here.

\begin{table}[htb]
\begin{center}
\begin{tabular}{ccccccccc}
& $N_\text{cut}$ & \quad \quad & &
$N_\text{cut}$ & \quad \quad & & $N_\text{cut}$ \\
$\TTWH$                        &  5.1 & &
  $t \bar t$                   &  0 & &
  $t \bar t b \bar b$ EW       &  0.4 \\
$\TTHH$                        &  3.5 & &
  $t \bar tj$                  &  5 & &
  $t \bar t c \bar c$ EW       &  0.0 \\
$\TTZH$                        &  1.8 & &
  $t \bar t2j$                 &  6 & &
  $Wjjjjjj$                    &  0 \\
$\TTWZ$                        &  0.4 & &
  $t \bar t3j$                 &  4  & &
  $W b \bar b jjjj$            &  0 \\
$\TTZZ$                        &  0.0   & &
  $t \bar t4j$                 &  2 & &
  $W c \bar c jjjj$            &  0 \\
$T \bar T b \bar b$            &  0.1 & &
  $t \bar t5j$                 &  0 & &
  $W b \bar b b \bar b$        &  0 \\
$t \bar t H$                   &  2.5 & &
  $t \bar t b \bar b$          &  4 & &
  $Zjjjjjj$                    &  0 \\
            &   & &
  $t \bar t c \bar c$          &  0 & &
  $Z b \bar b jjjj$            &  0 \\

   &   & &
   &   & & 
  $Z c \bar c jjjj$            &  0 \\

\end{tabular}
\caption{Analysis I: number of events $N_\text{cut}$ after the selection
criteria in Eqs.~(\ref{ec:cut11}).}
\label{tab:anI}
\end{center}
\end{table}

We remark that the signal itself has additional
higher order contributions $\ttHnj$, with $n \geq 1$, which have not been
included in the same way as $\ttnj$ because the implementation of the
matching prescription is not yet available (and also for consistency with the
calculation of $T \bar T$, in which only the lowest order $n=0$ can be
generated).
When higher order processes are included, there are two alternatives for the
likelihood analysis: (i) keep using the $\mathcal{N}_\text{jet}$ distribution
for $t \bar
t H$ in Fig.~\ref{fig:anI-vars1}, which suppresses $\ttnj$ but also $\ttHnj$ for
larger $n$; (ii) use a new $\mathcal{N}_\text{jet}$ distribution for $\ttHnj$,
which may
improve the results. The first option can always be followed, and will of course
lead to better results than the ones shown here (this is the reason why we have
not included any $\kfac$ factors in the signals). Thus, the results shown here
are conservative. From the number of $\ttnj$ events in Table~\ref{tab:anI} we
can estimate that the inclusion of higher $\ttHnj$ processes would double the
sensitivity at least. These comments also apply to the case in which
$\mathcal{N}_\text{jet}$ is not included in the likelihood but a cut on this
variable is performed (see the next section).

The new Higgs signals from $T \bar T$ decays enhance the observability of the
Higgs boson. Despite the very different kinematics of this process, and the fact
that the reconstruction is aimed at identifying $t \bar t H$ production,
$T \bar T$ events are more signal- than background-like, as it 
can be observed in Figs.~\ref{fig:anI-vars1}-- \ref{fig:anI-vars2}.
Hence, they are not very
suppressed by the kinematical cuts, and enhance the Higgs sensitivity
by a factor of 6, $\mathcal{S}_{20} = 2.03\sigma$. This
improvement is sufficient to have hints of the Higgs boson with a luminosity
of 30 fb$^{-1}$. However, one can do much better with a dedicated
reconstruction
aiming to detect the new quark.

\subsection{Analysis II: $\boldsymbol{T \bar T}$ reconstruction}
\label{sec:4.2}

The three different $T \bar T$ decay channels considered yield final states
with four or six $b$ quarks, and lead to signal events with four, five and six
or more $b$-tagged jets.
(Due to mistags, the number of $b$ jets may be occasionally
larger than the number of $b$ quarks at the partonic level.)
The number of events corresponding to each decay channel and number of $b$ jets
are collected in Table~\ref{tab:btag}, including also the SM background.

\begin{table}[htb]
\begin{center}
\begin{tabular}{lcccc}
           & Total   & 4 tags  & 5 tags & $\geq 6$ tags \\
$\TTWH$    & 339.0   & 303.2   &  33.7  &  2.1 \\
$\TTHH$    & 262.7   & 166.0   &  76.5  & 20.2 \\
$\TTZH$    & 130.5   & 97.9    &  27.3  &  5.3 \\
Background & 13158.9 & 12572.4 & 561.1  & 25.4
\end{tabular}
\end{center}
\caption{Analysis II: Number of events (for 30 fb$^{-1}$) with four,
five and six or more
$b$ tags, for each of the signal processes and the SM background.}
\label{tab:btag}
\end{table}

The discovery potential is higher if signal and background samples are
separated according to their $b$ jet multiplicity. This is also convenient
from the point of view of the signal reconstruction.
 The two main signal channels,
\begin{align}
& T \bar T \to W^+ b \, H \bar t / H t \, W^- \bar b 
 \to W^+ b W^- \bar b H \to f_1 \bar f_1' b \bar f_2 f_2' \bar b
 \;  b \bar b   & (WH) \,, \notag \\
& T \bar T \to H t \, H \bar t 
 \to W^+ b W^- \bar b H H \to f_1 \bar f_1' b \bar f_2 f_2' \bar b
 \;  b \bar b \;  b \bar b & (HH) \,,
\label{ec:rec12}
\end{align}
have four and six $b$ quarks in the final state, respectively, and different
kinematics. Hence, for the reconstruction the events are classified as follows:
\begin{itemize}
\item Events with four $b$ tags are assigned to the $WH$ mode and reconstructed
accordingly.
\item Events with six or more $b$ tags are assigned to the $HH$ mode.
\item Events with five tags are assumed to belong to the $HH$ mode as well if
there are at least three non-$b$ jets (the sixth $b$ jet is taken to be one of
the non-tagged ones). A small fraction $\simeq 5$\% which only has two light
jets is reconstructed as in the $WH$ mode.
\end{itemize}
This separation allows for a better reconstruction of
$\TTHH$ events with five or more $b$ tags, which amount to 36.6\% of this
channel and have a much smaller background.
The remaining $\TTHH$ events only have four tags, and they are
reconstructed as in the $\TTWH$ channel.\footnote{We have also tried a
reconstruction of the $HH$ channel with only four $b$ jets.
This requires taking two light jets (among the many ones present in general)
as if they were $b$ jets, with a minimum of 2160 combinations (for a minimum
of four light jets) for the reconstruction.
For events with four $b$ jets, we have thus tried a mixed procedure, 
selecting the channel which best fits the event kinematics.
This improves
the mass distributions for the $\TTHH$ signal but slightly degrades them for
the $\TTWH$ channel and concentrates the background in the
region of interest $M_H^\mathrm{rec} = 100-140$ GeV, giving worse results.}
In both methods the heavy quark mass is not used in order to not
bias the SM background towards this invariant mass value.
The reconstruction is done by trying all
possible pairings for light and $b$ jets, and selecting the one which
best resembles the kinematics of the decay channel considered.

\subsubsection{$4b$ final states}

\begin{figure}[t]
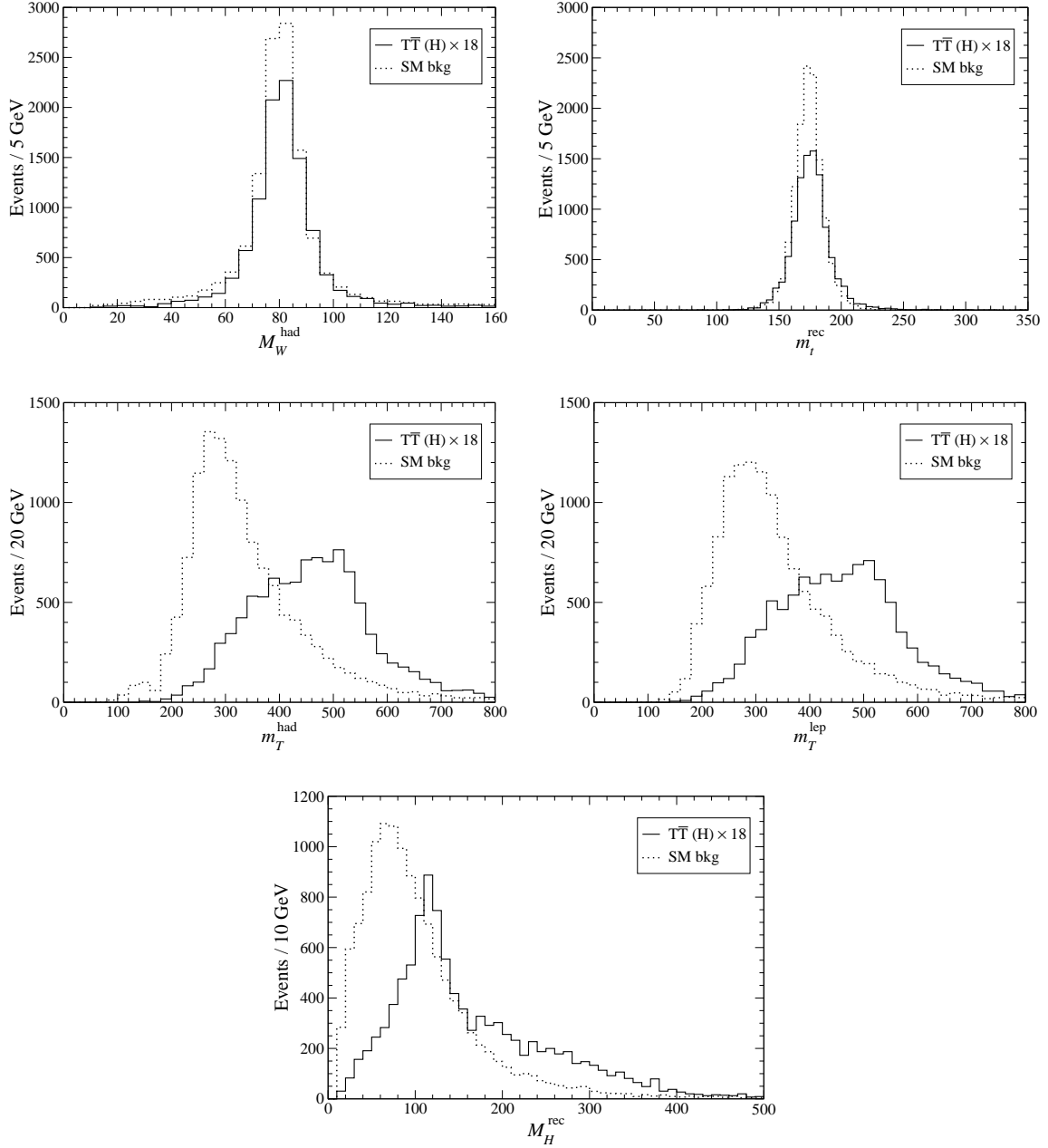

\begin{center}
\begin{tabular}{cc}
\epsfig{file=Figs/mWhad-anII4b.eps,height=5.2cm,clip=}
   &  \epsfig{file=Figs/mtrec-anII4b.eps,height=5.2cm,clip=}  \\[0.4cm]
\epsfig{file=Figs/mThad-anII4b.eps,height=5.2cm,clip=}
   & \epsfig{file=Figs/mTlep-anII4b.eps,height=5.2cm,clip=} \\[0.4cm]
\multicolumn{2}{c}{\epsfig{file=Figs/mH-anII4b.eps,height=5.2cm,clip=}}
\end{tabular}
\caption{Analysis II ($4b$ final states):
Reconstructed masses of the
hadronic $W$, the top quark, the hadronic and leptonic heavy quarks and the
Higgs boson, for the background and the sum of $T \bar T$ Higgs signals.}
\label{fig:anII4b-rec}
\end{center}
\end{figure}

We reconstruct the hadronic $W$ boson from a pair of light jets
$j_1$ and $j_2$, and the leptonic $W$ from the charged lepton and missing
transverse momentum. 
With the $W$ momenta determined up to a twofold ambiguity, we identify the
two $b$ quarks $b_T$ and $b_t$ coming from the decays $T \to Wb$, $t \to Wb$.
There are 24 possibilities for the pairing,
because: (i) the heavy quark decaying to $W b$ (irrespectively of whether it is
$T$ or $\bar T$) may have the $W$ boson decaying hadronically or leptonically;
(ii) the quark $b_T$ may correspond to each one of the four $b$-tagged jets
in the final state, and the three remaining ones are then produced in the
cascade decay $T \to H t \to b \bar b W b$; (iii) the quark $b_t$ from
the top decay can be any of the latter three.
Among the 48 resulting possibilities (plus different choices of $j_1$ and
$j_2$), we select the one minimising the quantity
\begin{equation}
\Delta m^2_{WH} = \frac{(\mThad-\mTlep)^2}{S_T^2}
+ \frac{(\mtrec-m_t)^2}{S_t^2} + \frac{(\mwhad-M_W)^2}{S_W^2} \,,
\end{equation}
where $\mtrec$ corresponds to the intermediate top quark (which may decay
hadronically or leptonically), and
$\mThad$, $\mTlep$ are the reconstructed masses of the hadronic and leptonic
$T$ quarks (independently of whether they decay to $Wb$ or $Ht$).
$S_T$, $S_t$ and $S_W$ are taken as $S_T = 100$ GeV,
$S_t = 20$ GeV and $S_W = 10$ GeV. 
No cuts are applied at this level.
For the best pairing, the two remaining $b$ jets not assigned to
the $T$ and $t$ decays correspond to
the Higgs boson. The reconstructed masses are shown
in Fig.~\ref{fig:anII4b-rec} for the sum of signal channels and the SM
background.

\begin{figure}[p]
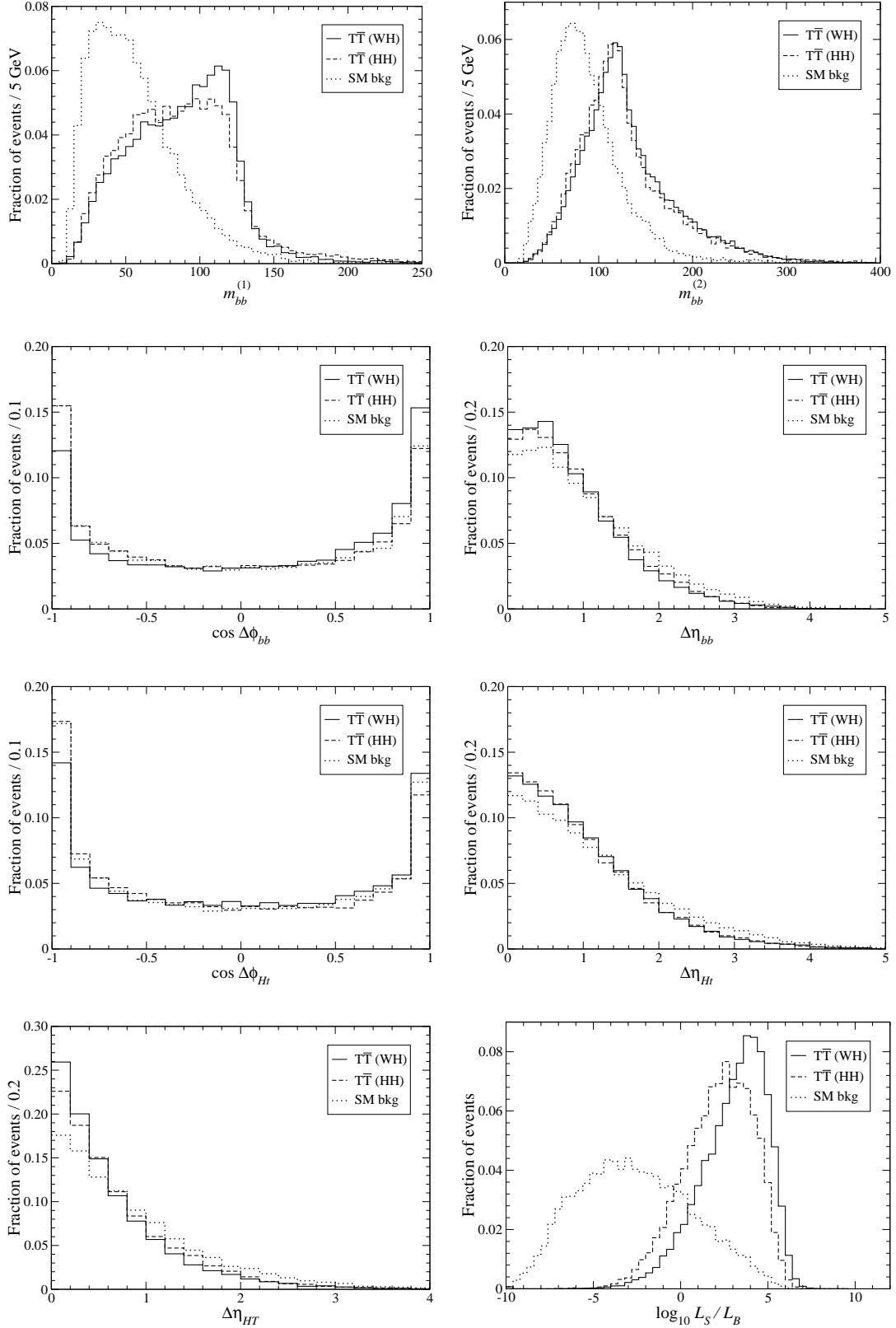

\begin{center}
\begin{tabular}{cc}
\epsfig{file=Figs/mbb1_n-anII4b.eps,height=4.9cm,clip=}
   &  \epsfig{file=Figs/mbb2_n-anII4b.eps,height=4.9cm,clip=} \\[0.4cm]
\epsfig{file=Figs/cPhbb_n-anII4b.eps,height=4.9cm,clip=}
   &  \epsfig{file=Figs/dhbb_n-anII4b.eps,height=4.9cm,clip=} \\[0.4cm]
\epsfig{file=Figs/cPhHtop_n-anII4b.eps,height=4.9cm,clip=}
   &  \epsfig{file=Figs/dhHtop_n-anII4b.eps,height=4.9cm,clip=} \\[0.4cm]
\epsfig{file=Figs/dhHt_n-anII4b.eps,height=4.9cm,clip=} 
  &  \epsfig{file=Figs/logLSf_n-anII4b.eps,height=4.9cm,clip=}
\end{tabular}
\caption{Analysis II ($4b$ final states): Normalised variables
$m_{bb}^{(1)}$, $m_{bb}^{(2)}$, $\cos \Delta \phi_{bb}$, $\Delta \eta_{bb}$,
$\cos \Delta \phi_{Ht}$, $\Delta \eta_{Ht}$ and $\Delta \eta_{HT}$
(defined in the text), used in the likelihood analysis. Log-likelihood function
$\log_{10} L_S / L_B$.}
\label{fig:anII4b-vars2}
\end{center}
\end{figure}

We build signal and background likelihood functions using:
\begin{itemize}
\item The reconstructed masses $\mThad$, $\mTlep$.
\item Variables characterising the high transverse momentum of the signal:
the total transverse energy $H_T$, the missing energy
$\ptmiss$, the maximum and second maximum $p_t$ of the $b$ jets $\ptbmax$ and
$\ptbmaxx$, and the second maximum $p_t$ of the light jets $\ptlmaxx$.
\item The energy of the charged lepton in the heavy quark rest frame,
$E_\ell^*$. This distribution has a long tail for $\TTWH$ signal events, not
only because of the large $T$ mass but also due to spin effects \cite{top06}.
\item The smallest invariant mass of a $bb$ pair $m_{bb}^{(1)}$ and the second
smallest one $m_{bb}^{(2)}$.
\item Angular quantities characterising the topology of the event: the azimuthal
angle and rapidity difference (i) between the two $b$ jets assigned
to the Higgs, $\Delta \phi_{bb}$ and $\Delta \eta_{b b}$;
(ii) between the Higgs and the reconstructed top quark,
$\Delta \phi_{Ht}$ and $\Delta \eta_{H t}$; (iii)
between the Higgs and its parent $T$ quark, $\Delta \eta_{HT}$.
\end{itemize}
The distributions of these variables are presented in
Figs.~\ref{fig:anII4b-vars1}, \ref{fig:anII4b-vars2}. We remark again that the
selection of variables is not arbitrary, and some variables not considered,
{\em e.g.} the transverse momentum of the charged lepton or the maximum
transverse momentum of the light jets, have not been included
because they actually reduce the discriminating power with respect to the set
of variables above. This surprising fact is due to the correlation among
variables, and is further explained in the appendix.
We distinguish three likelihood classes: the $\TTWH$ and $\TTHH$ signals and
the background. The signal likelihood is defined as 
the sum of the likelihoods of the two signal classes,
$L_S = L_{S_1} + L_{S_2}$.
The logarithm of $L_S / L_B$ is plotted in Fig.~\ref{fig:anII4b-vars2}.
We observe that the $\TTWH$ distributions are in general more distinguishable
from the background than the $\TTHH$ ones. This results in a cleaner separation
between $\TTWH$ and the background.

\begin{figure}[p]
\begin{center}
\begin{tabular}{cc}
\epsfig{file=Figs/mThad_n-anII4b.eps,height=4.9cm,clip=}
   & \epsfig{file=Figs/mTlep_n-anII4b.eps,height=4.9cm,clip=} \\[0.4cm]
\epsfig{file=Figs/HT_n-anII4b.eps,height=4.9cm,clip=}
   &  \epsfig{file=Figs/ptmiss_n-anII4b.eps,height=4.9cm,clip=} \\[0.4cm]
\epsfig{file=Figs/ptbmax_n-anII4b.eps,height=4.9cm,clip=}
   & \epsfig{file=Figs/ptbmax2_n-anII4b.eps,height=4.9cm,clip=} \\[0.4cm]
\epsfig{file=Figs/ptlmax2_n-anII4b.eps,height=4.9cm,clip=} 
   & \epsfig{file=Figs/Elep_n-anII4b.eps,height=4.9cm,clip=} 
\end{tabular}
\caption{Analysis II ($4b$ final states): Normalised variables
$\mThad$, $\mTlep$, $H_T$, $\ptmiss$, $\ptbmax$, $\ptbmaxx$, $\ptlmaxx$
and $E_\ell^*$
(defined in the text), used in the likelihood analysis.}
\label{fig:anII4b-vars1}
\end{center}
\end{figure}

For event selection we again use a hybrid method, with cuts on reconstructed
masses, jet multiplicity and signal likelihood. The selection criteria are
\begin{align}
& \log_{10} L_S/L_B \geq 3.9 \,, \notag \\
& \mathcal{N}_\text{jet} \leq 7 \,, \notag \\
& 100 ~\text{GeV} \leq \mhrec \leq 140 ~\text{GeV} \,, \notag \\
& 350~\text{GeV} \leq \mThad,\mTlep \leq 650~\text{GeV} \,.
\label{ec:sel4b}
\end{align}
The numbers of events after these cuts are collected in
Table \ref{tab:anII4b}. The $\ttnj$ background with larger $n$ has larger
transverse momenta and is less affected by the cut on likelihood, but it is
suppressed by the cut on jet multiplicity. $W/Z$ plus jets is insignificant.
We also note the smaller efficiency for the $\TTHH$ signal, expected since its
likelihood function has a larger overlap with the background, see
Fig.~\ref{fig:anII4b-vars2}. Additionally, $\TTHH$ decays with four $b$-tagged
jets have a larger light jet multiplicity, and are more affected by the
requirement $\mathcal{N}_\text{jet} \leq 7$. The same comments made in the
preceding subsection regarding the cut on $\mathcal{N}_\text{jet}$ and higher
order signal processes apply here.

Before calculating the statistical significance of the Higgs signals from
$T \bar T$ decays it is important to draw attention
to the fact that, since neither
the $T$ quark nor the Higgs boson have been discovered at present, there are two
possible definitions for what we consider as signal
and background. The first one would be to take as background just the SM
processes in Table~\ref{tab:ang} (excluding $t \bar t H$), and for the signal
$t \bar t H$, $T \bar T$ (in all decay modes) and $T \bar T b \bar b$.
The second possibility is to take as background the SM processes
(slightly modified by the presence of the heavy quark) plus $T \bar T \,
(WZ,ZZ)$ and $T \bar T b \bar b$ {\em in the absence} of a Higgs boson
(see Table \ref{tab:ang}). Signal plus background is then constituted by the SM
processes,
plus $T \bar T \, (WZ,ZZ)$ and $T \bar T b \bar b$ {\em with} a Higgs boson,
and Higgs production processes $t \bar t H$ and $T \bar T \, (WH,HH,ZH)$.
The ``signal'', that is, the excess of events over the background, is thus
$t \bar t H$ plus $T \bar T \, (WH,HH,ZH)$ plus the difference
between $T \bar T \, (WZ,ZZ)$ and $T \bar T b \bar b$ with and without a Higgs
boson, that is,
\begin{eqnarray}
B & = & \text{SM bkg.} + T \bar T (WZ,ZZ;\noH) \,, \notag \\
S & = & t \bar t H (T) + T \bar T (WH,HH,ZH) \notag \\
& &  + \left[T \bar T (WZ,ZZ)
- T \bar T (WZ,ZZ;\noH) \right] + \Delta \, \text{SM bkg.}
\label{ec:delta}
\end{eqnarray}
The term in brackets is always negative, and the difference in SM background is
negligible.
Both conventions lead to appreciably different results, and we adopt the
latter, which is more conservative. (This amounts to considering that the $T$
quark will have been discovered before the Higgs boson.)
With this definition, 
the signal significance is $\mathcal{S}_{20} = 6.43\sigma$, including a
20\% systematic error.

\begin{table}[htb]
\begin{center}
\begin{tabular}{ccccccccc}
& $N_\text{cut}$ & \quad \quad & &
$N_\text{cut}$ & \quad \quad & & $N_\text{cut}$ \\
$\TTWH$                        &  36.2 & &
  $t \bar t$                   &  1 & &
  $t \bar t c \bar c$ EW       &  0.0 \\
$\TTHH$                        &  5.4 & &
  $t \bar tj$                  &  0 & &
  $t \bar t c \bar c$ EW ($T$) &  0.0 \\
$\TTZH$                        &  2.9 & &
  $t \bar t2j$                 &  2 & &
  $Wjjjjjj$                    &  0 \\
$\TTWZ$                        &  0.8 & &
  $t \bar t3j$                 &  3  & &
  $W b \bar b jjjj$            &  0 \\
$\TTWZ$  ($\noH$)              &  2.1 & &
  $t \bar t4j$                 &  8 & &
  $W c \bar c jjjj$            &  0 \\
$\TTZZ$                        &  0.0   & &
  $t \bar t5j$                 &  2 & &
  $W b \bar b b \bar b$        &  0 \\
$\TTZZ$  ($\noH$)              &  0.1 & &
  $t \bar t b \bar b$          &  3 & &
  $Zjjjjjj$                    &  0 \\
$T \bar T b \bar b$            &  0.0 & &
  $t \bar t c \bar c$          &  2 & &
  $Z b \bar b jjjj$            &  0 \\

$T \bar T b \bar b$ ($\noH$)   &  0.1 & &
  $t \bar t b \bar b$ EW       &  0.5 & & 
  $Z c \bar c jjjj$            &  0 \\
$t \bar t H$ ($T$)             &  0.8 & &
  $t \bar t b \bar b$ EW ($T$) &  0.3 \\

\end{tabular}
\caption{Analysis II ($4b$ final states):
Number of events (for 30 fb$^{-1}$) after the kinematical cuts in
Eq.~(\ref{ec:sel4b}).}
\label{tab:anII4b}
\end{center}
\end{table}

\subsubsection{$5b$ and $6b$ final states}

\begin{figure}[t]
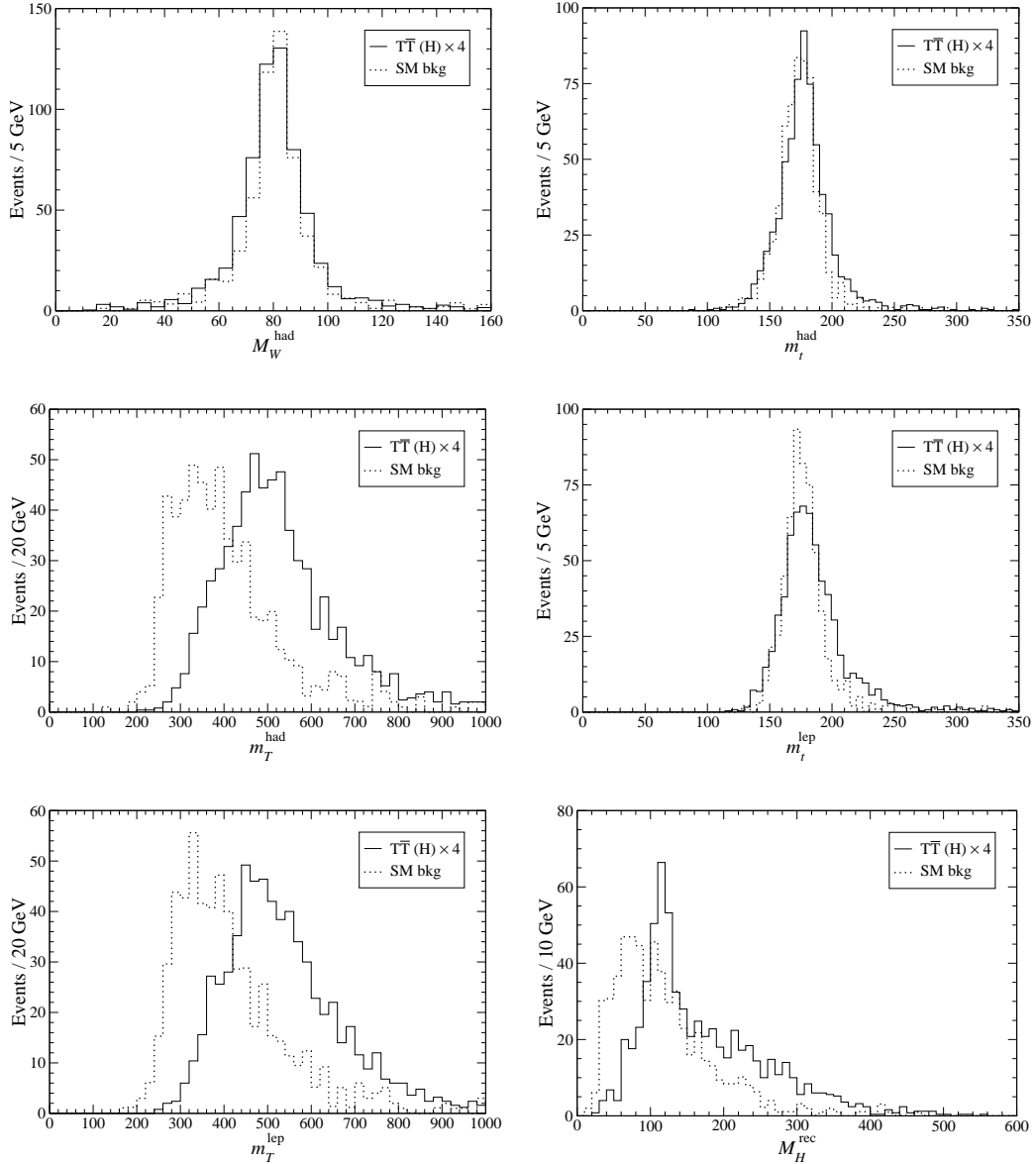

\begin{center}
\begin{tabular}{cc}
\epsfig{file=Figs/mWhad-anII5b.eps,height=4.8cm,clip=}
   &  \epsfig{file=Figs/mtrec-anII5b.eps,height=4.8cm,clip=}  \\[0.4cm]
\epsfig{file=Figs/mThad-anII5b.eps,height=4.8cm,clip=}
   &  \epsfig{file=Figs/mtrec2-anII5b.eps,height=4.8cm,clip=}  \\[0.4cm]
\epsfig{file=Figs/mTlep-anII5b.eps,height=4.8cm,clip=}
   & \epsfig{file=Figs/mH-anII5b.eps,height=4.8cm,clip=}
\end{tabular}
\caption{Analysis II ($5b$, $6b$ final states):
Reconstructed masses of the
hadronic $W$, the hadronic and leptonic top and heavy quarks and the
Higgs boson, for the background and the sum of $T \bar T$ Higgs signals.}
\label{fig:anII-51}
\end{center}
\end{figure}

Reconstructing the decay $T \bar T \to Ht H \bar t \to H W^+ b H W^- \bar b$
requires identifying six $b$ jets in the final state. In the case of five $b$
tags, a light jet $j_b$ (if there are at least three) may
be assumed to come from a $b$ quark as well.
The hadronic $W$ boson is reconstructed from a pair of untagged jets
$j_1$ and $j_2$.  The leptonic $W$ is reconstructed
from the charged lepton momentum and missing energy.
Each $W$ boson is associated to three $b$ jets to reconstruct the
momenta of the $T$ quarks (there are 20 combinations). For each choice, there
are $3 \times 3$ possibilities to associate two $b$ jets to the hadronic and
leptonic $W$, in order to reconstruct the two top quarks. The two remaining
pairs of $b$
jets $(b_1,b_2)$, $(b_3,b_4)$ are assumed to come from the decays of the two
Higgs bosons, with reconstructed masses $M_{H_1}^\mathrm{rec} = m_{b_1 b_2}$
(associated to the hadronic top),
$M_{H_2}^\mathrm{rec} = m_{b_3 b_4}$ (associated to the leptonic one).
Among the 360 resulting possibilities (plus different choices of $j_1$, $j_2$
and $j_b$), we select the one minimising the quantity
\begin{eqnarray}
\Delta m^2_{HH} & = &
\frac{(\mThad-\mTlep)^2}{S_T^2} + 
\frac{(M_{H_1}^\mathrm{rec}-M_{H_2}^\mathrm{rec})^2}{S_H^2} +
 \frac{(\mthad-m_t)^2}{S_t^2}
+ \frac{(\mtlep-m_t)^2}{S_t^2} \notag \\
& & + \frac{(\mwhad-M_W)^2}{S_W^2} \,.
\end{eqnarray}
We take $S_T = 100$ GeV, $S_t = 20$ GeV, $S_W = S_H = 10$ GeV. 
No cuts are applied at this level.
The reconstructed masses are shown
in Fig.~\ref{fig:anII-51} for the sum of the signal channels and the SM
background. We define the reconstructed Higgs mass as the average of
$M_{H_1}^\text{rec}$ and $M_{H_2}^\text{rec}$. In this way, a sharper peak is
obtained.

\begin{figure}[p]
\begin{center}
\begin{tabular}{cc}
\epsfig{file=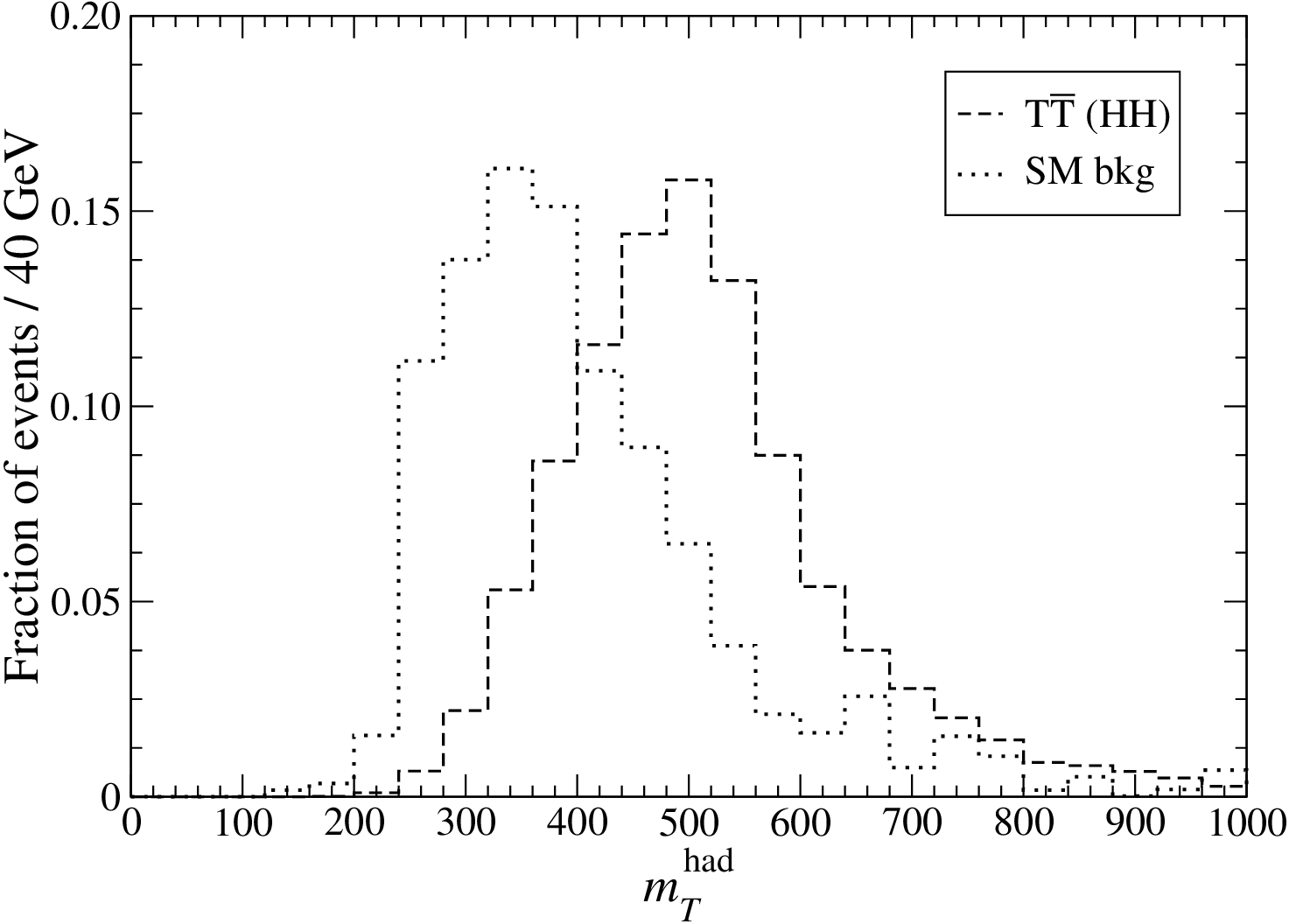,height=4.8cm,clip=}
  &  \epsfig{file=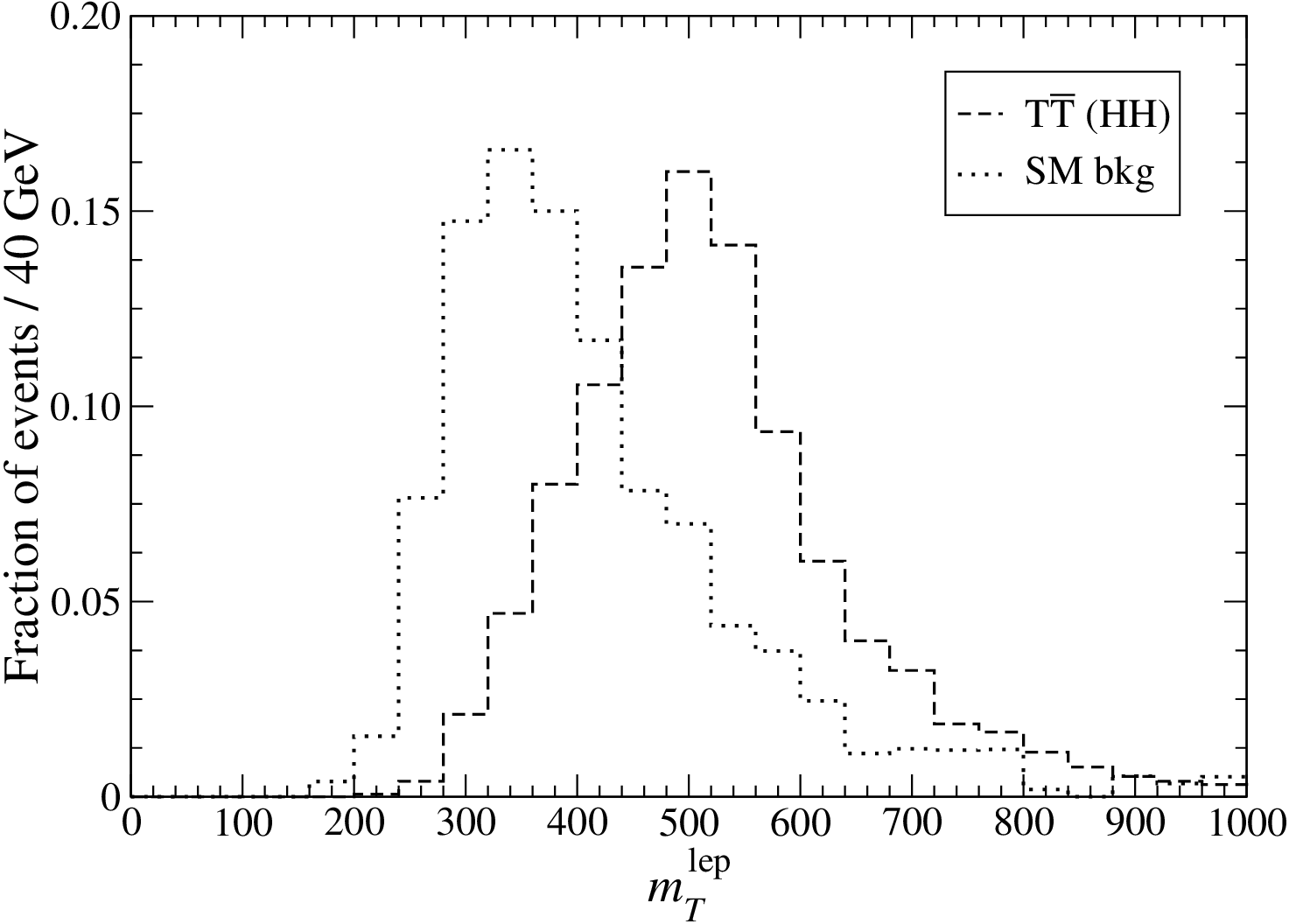,height=4.8cm,clip=} \\[0.4cm]
\epsfig{file=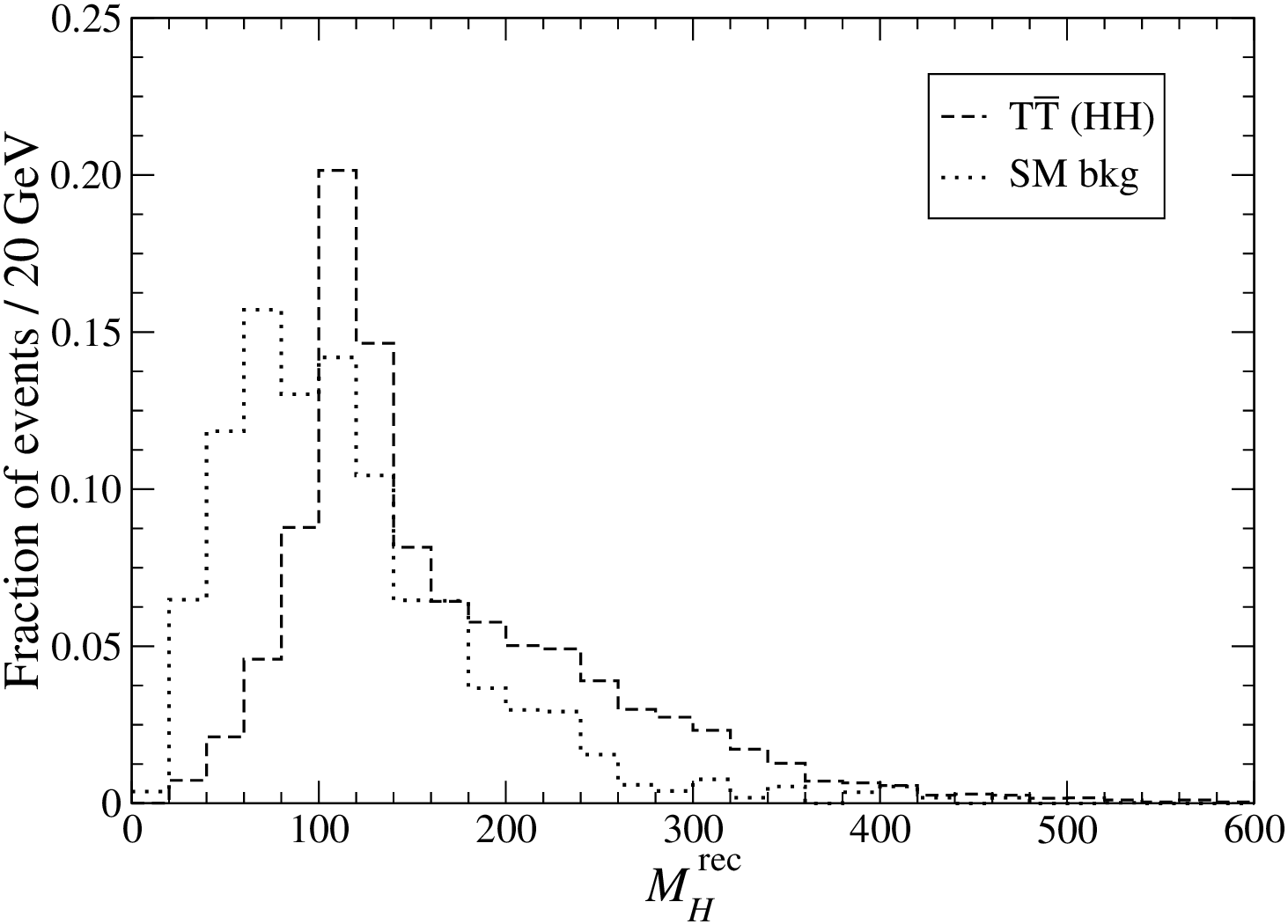,height=4.8cm,clip=}
  & \epsfig{file=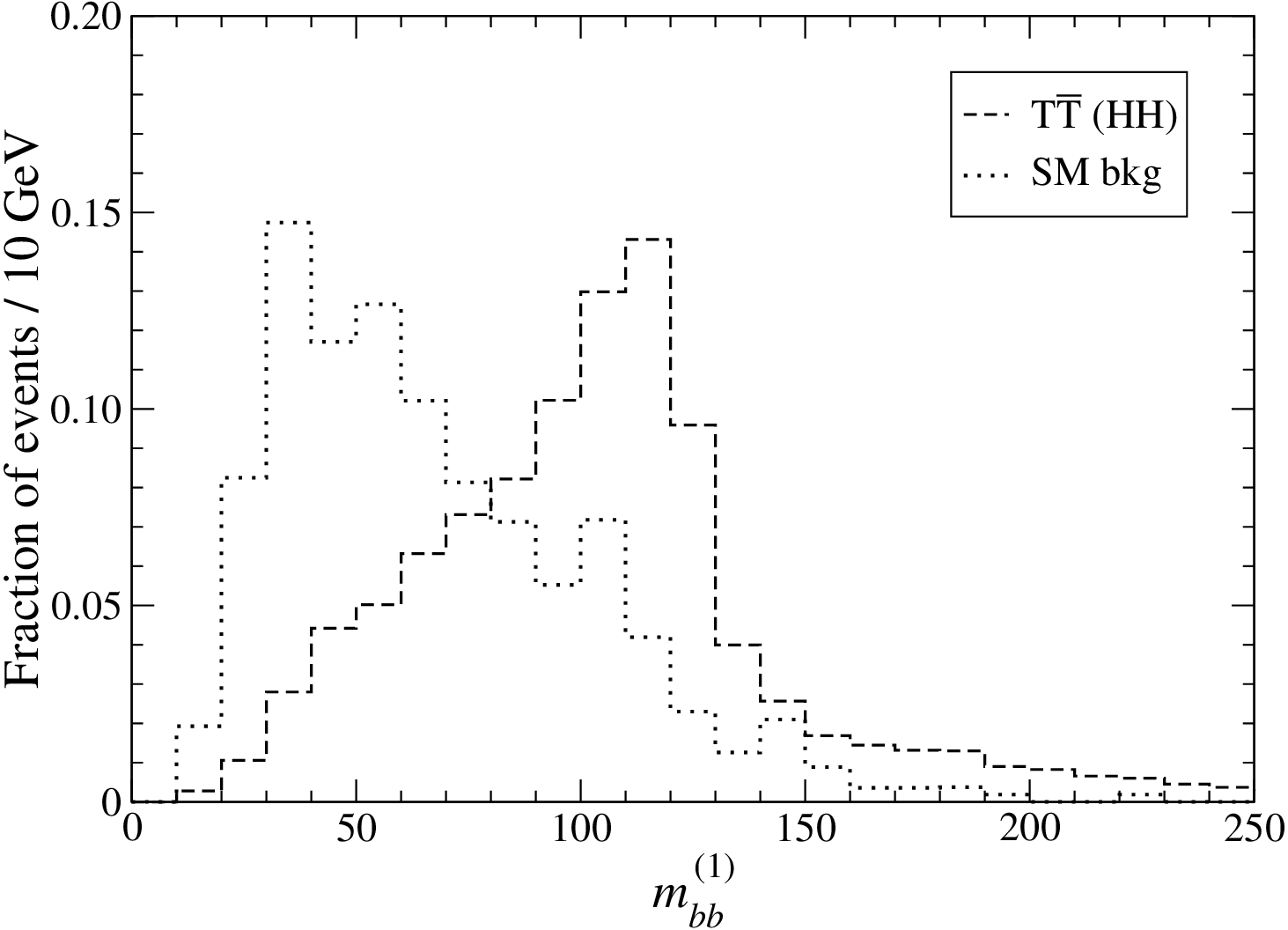,height=4.8cm,clip=} \\[0.4cm]
\epsfig{file=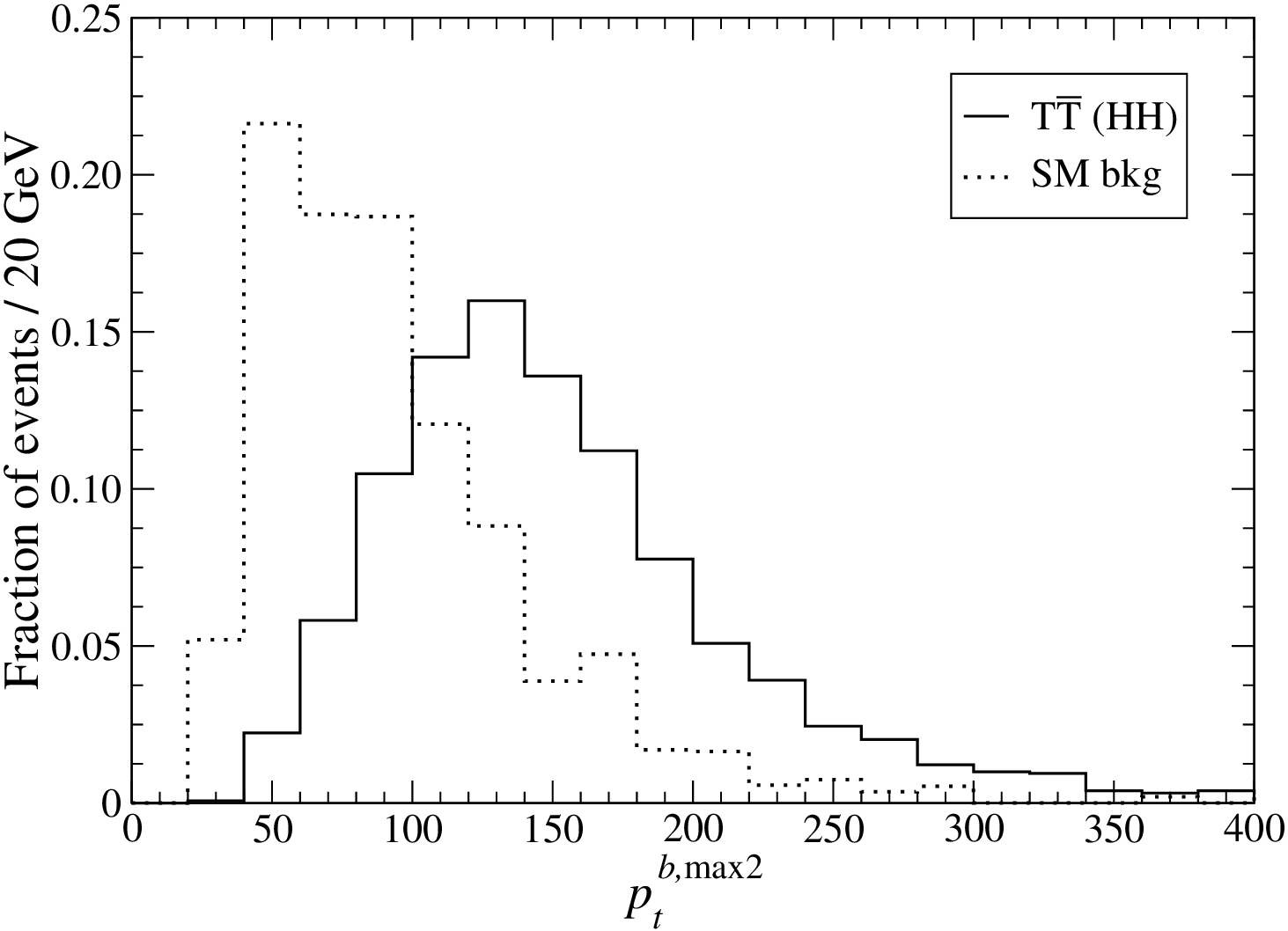,height=4.8cm,clip=}
  &  \epsfig{file=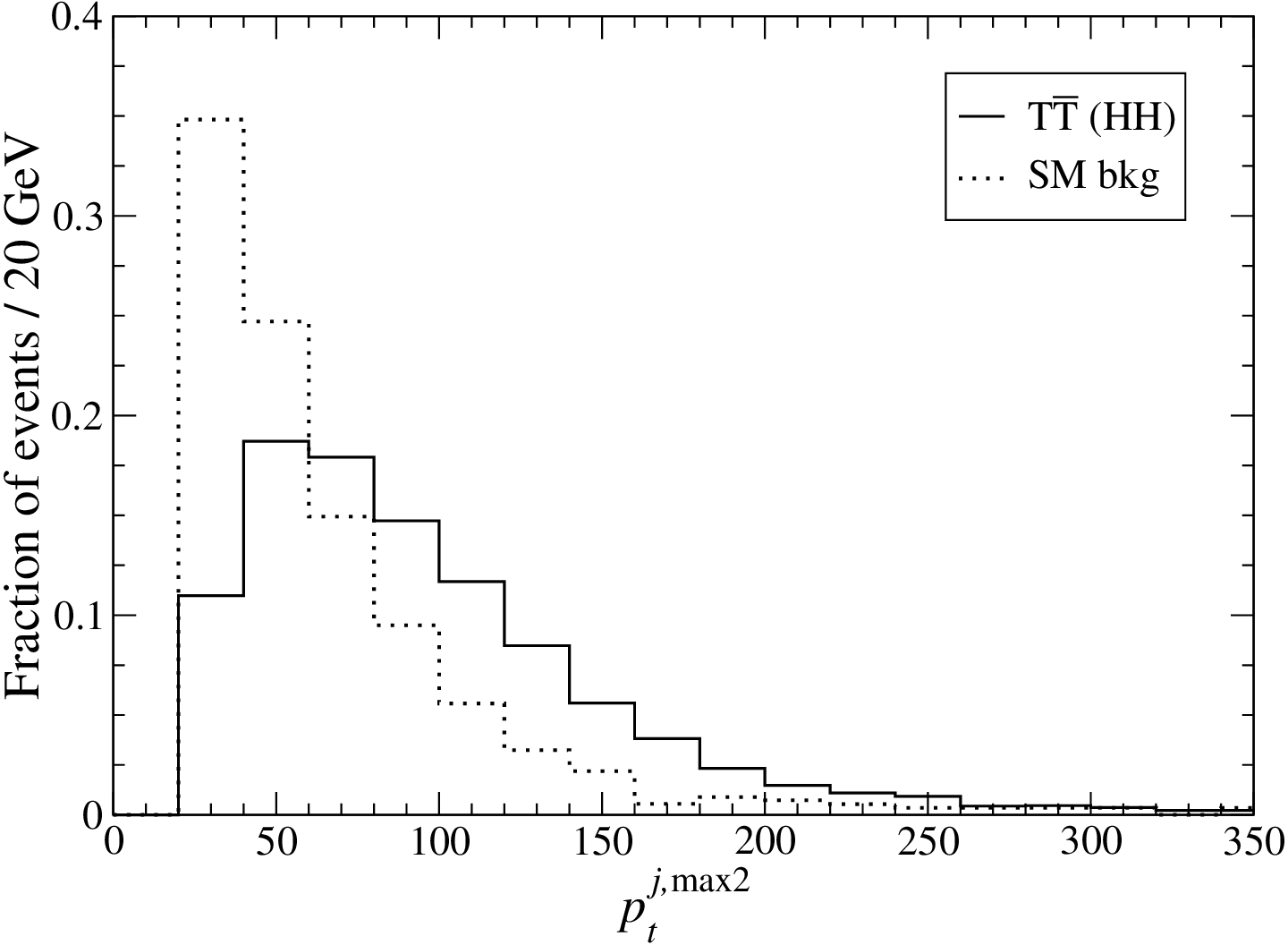,height=4.8cm,clip=} \\[0.4cm]
\epsfig{file=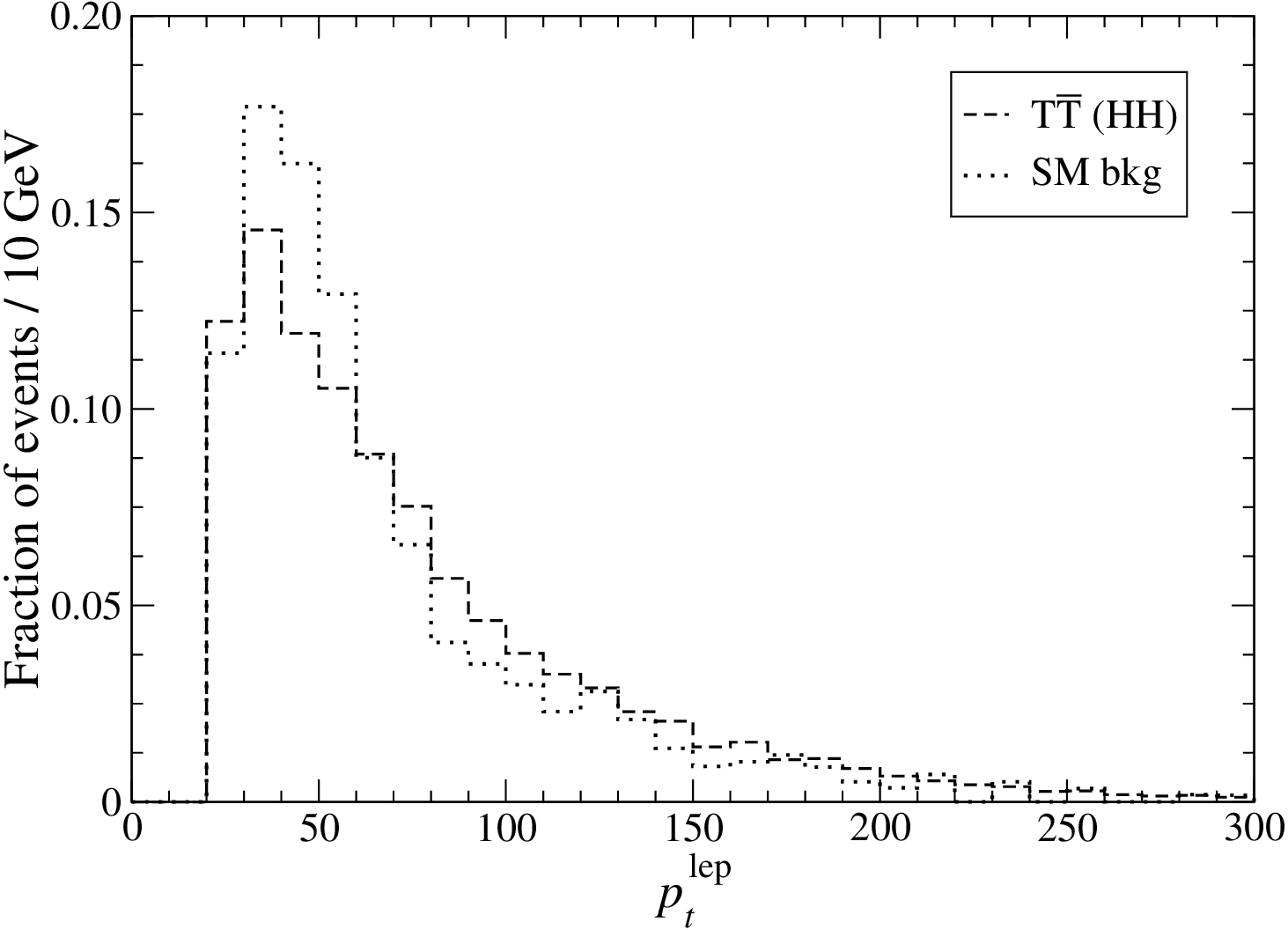,height=4.8cm,clip=}
  &  \epsfig{file=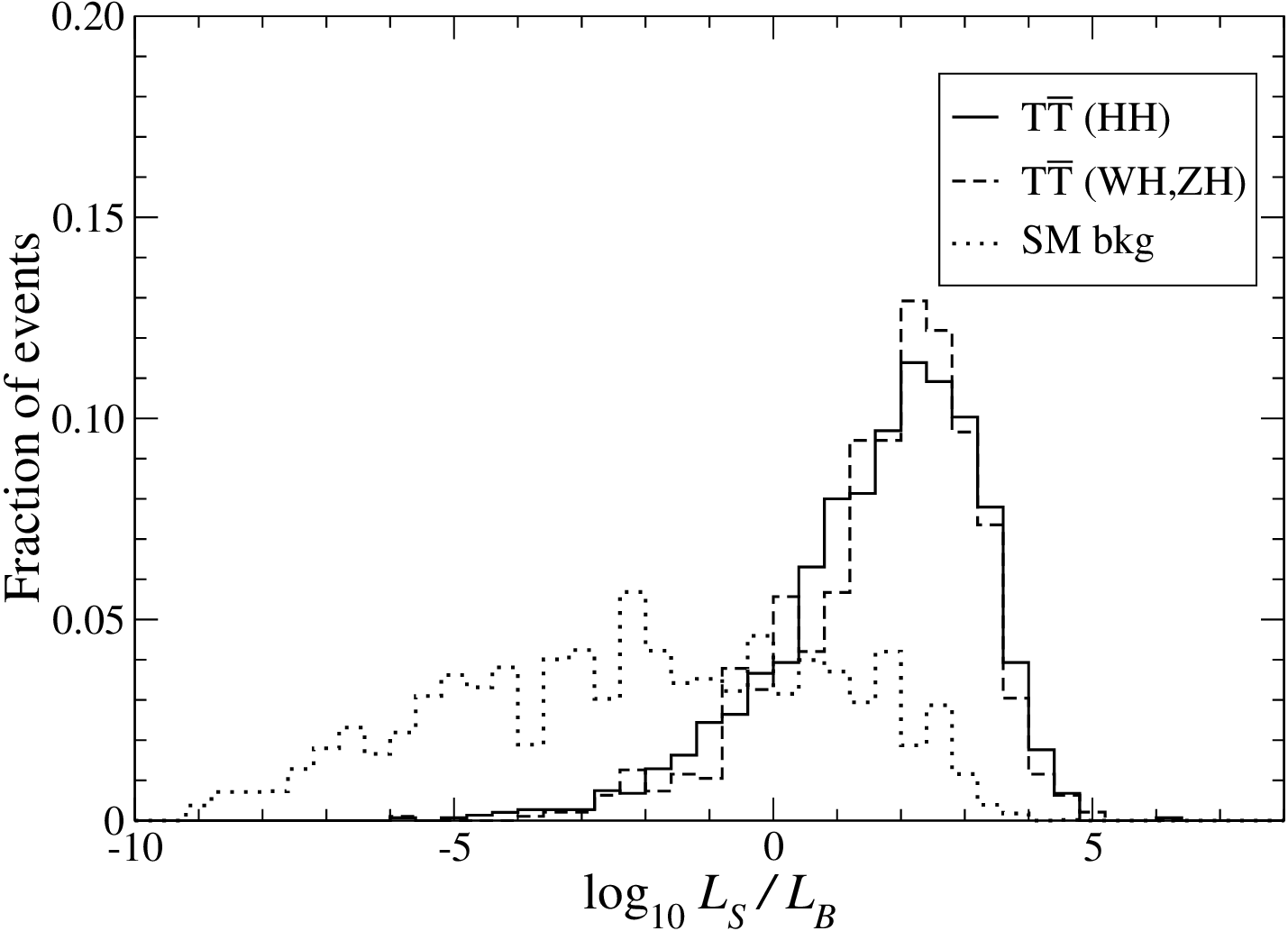,height=4.8cm,clip=}
\end{tabular}
\caption{Analysis II ($5b$, $6b$ final states): Normalised variables
$\mThad$, $\mTlep$, $\mhrec$, $m_{bb}^{(1)}$, $\ptbmaxx$, $\ptlmaxx$ and
 $\ptlep$, used in the likelihood
analysis. Log-likelihood function.} 
\label{fig:anII5b-vars1}
\end{center}
\end{figure}

In these final states the SM background is already very small, and performing
kinematical cuts on reconstructed Higgs and heavy quark masses or light jet
multiplicity
can easily reduce the signal significance. Therefore, for this analysis
we include these variables in the likelihood functions, and only perform loose
cuts on the signal likelihood.
The variables used are $\mThad$, $\mTlep$, 
$\mhrec$, $m_{bb}^{(1)}$, $H_T$, $\ptbmax$, $\ptbmaxx$, and $\ptlmaxx$,
defined in the previous subsection,  the jet multiplicity and the 
the charged lepton transverse momentum $\ptlep$. We only use two classes,
for the $\TTHH$ signal and the background, and 
the same distributions are used for final states with 5 and 6 $b$ quarks.
The normalised variables are presented in Fig.~\ref{fig:anII5b-vars1} except
$H_T$ and $\ptbmax$ which are very similar to the plots
in Fig.~\ref{fig:anII4b-vars1} and the jet multiplicity, shown in
Fig.~\ref{fig:anI-vars2}. The log-likelihood function is also presented in
Fig.~\ref{fig:anII5b-vars1}. We point out that the signal likelihood for the
$\TTWH$ and $\TTZH$ processes is very high even without using a separate
class for them.

We suppress the background by requiring
\begin{align}
& \log_{10} L_S/L_B \geq 2.6 & & (5b) \,, \notag \\
& \log_{10} L_S / L_B \geq 0 & & (6b) \,.
\label{ec:sel5b6b}
\end{align}
The number of events after these cuts are collected in Table~\ref{tab:anII5b6b}.
For 30 fb$^{-1}$ of luminosity, the statistical
significance of the Higgs signal is $\mathcal{S}_{20} = 6.02\sigma$,
$\mathcal{S}_{20} = 5.63\sigma$ for $5b$ and $6b$ final states, respectively.
We observe that the $t \bar t b \bar b$ background acquires increasing relevance
in these final states with five and six $b$-tagged jets. In order to
have a good estimate of the effect of higher order processes $t \bar tb \bar b
j$, $t \bar t b \bar b jj$, etc. we have included a
factor $\kfac=2.05$ into its tree-level cross section, as explained in
section~\ref{sec:3}. However, the kinematics of the higher order 
processes might be important and a detailed simulation (when a Monte Carlo
generator including a matching prescription for these processes is available)
is needed to confirm these results. Besides, we have explicitly
checked that the $t \bar t b \bar b b \bar b$ background, not included in our
simulations, is negligibly small.

\begin{table}[htb]
\begin{center}
\begin{tabular}{ccccccc}
&  $N_\text{cut}^{(5)}$ & $N_\text{cut}^{(6)}$ 
& \quad \quad & &
$N_\text{cut}^{(5)}$ & $N_\text{cut}^{(6)}$  \\
$\TTWH$                        &  13.3  & 2.0  & &
                     $t \bar t b \bar b$            &  8 & 7    \\
$\TTHH$                         &  23.4  & 17.6 & &
                     $t \bar t c \bar c$            &  0  & 0   \\
$\TTZH$                        &  8.1 & 4.6   & &
                     $t \bar t b \bar b$ EW         &  1.4 & 0.3   \\
$\TTWZ$                        &  1.6 & 0.4   & &
                     $t \bar t b \bar b$ EW ($T$)  &  0.9  & 0.5  \\
$\TTWZ$  ($\noH$)              &  3.5 & 0.7    & &
                     $t \bar t c \bar c$ EW          &  0.0 &  0.0   \\
$\TTZZ$                        &  0.6  & 0.2  & &
                     $t \bar t c \bar c$ EW ($T$)   &  0.0 & 0.0    \\
$\TTZZ$  ($\noH$)              &  1.2 & 0.6   & & 
                     $Wjjjjjj$                      &  0  & 0      \\
$T \bar T b \bar b$            &  0.1 & 0.0     & &
                     $W b \bar b jjjj$              &  2  & 1     \\
$T \bar T b \bar b$ ($\noH$)    &  0.1 & 0.0    & &
                     $W c \bar c jjjj$              &  0  & 0     \\
$t \bar t H$ ($T$)              &  1.4 & 0.3  & &
                     $W b \bar b b \bar b$          &  0  & 0    \\
$t \bar t$                      &  0  & 0     & &
                     $Zjjjjjj$                      &  0  & 0     \\
$t \bar tj$                     &  2  &  1     & &
                     $Z b \bar b jjjj$              &  0 & 0 \\
$t \bar t2j$                    &  0  & 0     & &
                     $Z c \bar c jjjj$               &  0 & 0 \\
$t \bar t3j$                    &  0  &  0     & &
                      \\
$t \bar t4j$                    &  1  &  0     & &
                      \\
$t \bar t5j$                    & 7 & 1   
\end{tabular}
\caption{Analysis II ($5b$, $6b$ final states):
Number of events (for 30 fb$^{-1}$) after the
selection cuts in Eqs.~(\ref{ec:sel5b6b})}
\label{tab:anII5b6b}
\end{center}
\end{table}

\subsubsection{Summary}

For a luminosity of 30 fb$^{-1}$, the statistical significances of the three
channels (including a 20\% background systematic uncertainty) are
\begin{align}
4b: & \quad S_{20} = 6.43 \sigma \,, \notag \\
5b: & \quad S_{20} = 6.02 \sigma \,, \notag \\
6b: & \quad S_{20} = 5.63 \sigma \,.
\end{align}
When the three channels are combined, a statistical significance of
$10.45\sigma$ is obtained for the $T \bar T$ Higgs signals.
This is a factor of
25 better than for $t \bar t H$ production, and offers a good opportunity to
quickly discover the Higgs boson in final states
containing a charged lepton and four or
more $b$ quarks.  Rescaling the expected signal and background rates (and using
Poisson statistics) it is found
that a $5\sigma$ discovery could be achieved approximately for 8 fb$^{-1}$.
This represents a reduction in luminosity by more than
one order of magnitude with respect to
$t \bar t H$ production in all $t \bar t$ decay channels,
and might be improved with less restrictive selection
cuts. This high sensitivity is due not only to the large $T \bar T$ cross
section, but also to the distinctive features of this signal, characterised by
large transverse momenta, high $b$ jet multiplicity and reconstructed invariant
masses peaking at $m_T$. At any rate, a likelihood analysis must be employed to
benefit from the distinctive kinematics and separate these signals from the
$\ttnj$ background, which also involves large transverse momenta for higher
values of $n$.

We finally comment on the experimental observation of the Higgs boson from $T
\bar T$ decays.
Although the reconstruction of the final state does not
explicitly make use of the new quark mass, the distributions used in the
probabilistic analysis do. Since the mass of an eventual heavy quark $T$ is
unknown, two alternatives are possible for the experimental search: (i) generate
sets of distributions and build likelihood functions for different values of
$m_T$ and compare them with real data; (ii) set generic kinematical cuts and
look for peaks in the invariant mass distributions. The second approach
gives sensitivities similar or worse than the ones obtained in this
section, and the analysis has been omitted for brevity. For illustration,
in the next section we will show how the new quark can be discovered with the
observation of peaks in the $\mThad$, $\mTlep$ distributions.

\section{Heavy quark discovery}
\label{sec:5}

Discovering the Higgs boson from $T \bar T$ decays implies the
discovery of the new quark. However, as emphasised in the paragraph before 
Eq.~(\ref{ec:delta}), the significances for the Higgs and $T$ quark discoveries
are different, due to the different classification of signals and backgrounds.
Using the data in
Tables~\ref{tab:anII4b}, \ref{tab:anII5b6b} and 
taking $t \bar t H$ as part of the background, the significances for $T$
discovery with 30 fb$^{-1}$ are
\begin{align}
4b: & \quad S_{20} = 6.93 \sigma \,, \notag \\
5b: & \quad S_{20} = 7.09 \sigma \,, \notag \\
6b: & \quad S_{20} = 6.28 \sigma \,,
\end{align}
with a combined significance $S_{20} = 11.74\sigma$. $5\sigma$ evidence of the
new quark (always assuming $m_T = 500$ GeV) could be achieved for
7 fb$^{-1}$.

It is also interesting to discover the new quark by observing peaks in
the $\mThad$, $\mTlep$ distributions. Quantifying the confidence level of such
peaks, so as to claim discovery, requires an appropriate background
normalisation. The procedure used here follows and extends
the one proposed in
Ref.~\cite{baur} for detecting anomalous couplings. 
Performing a $\chi^2$ fit to the binned data, a background
rescaling factor $\kappa$ can be obtained by minimising the quantity
\begin{equation}
\chi^2 = \sum_i \frac{(N_i - \kappa B_i)^2}{\kappa B_i} \,,
\end{equation}
where $i$ sums over the bins, $N_i$ are the numbers of events observed and
$B_i$ the expected background. The minimum is found for
\begin{equation}
\kappa^2 = \frac{1}{B} \sum_i \frac{N_i^2}{B_i} \,, 
\end{equation}
where $B = \sum_i B_i$ is the total expected background.
Since in a real experiment the number of events observed will include not
only the background but also a part from the signal itself, in most cases
$\kappa > 1$ will be found.
The uncertainty in this normalisation factor is given by
\begin{equation}
\delta \kappa^2 = \left[ \frac{3 \kappa^4}{B} + \frac{1}{B^2} \sum_i
\frac{N_i^4}{B_i^3} \right]^{1/2} \,.
\end{equation}
For a single bin we have $\kappa = N/B$, $\delta \kappa/\kappa = 1/\sqrt B$,
as expected. The statistical significance of the signal at the peak is
\begin{equation}
\mathcal{S}_\kappa \equiv S' / \sqrt{\kappa B + (\delta \kappa B)^2} \,,
\end{equation}
where $S' < S$ is the excess of events over the rescaled background. The second term
in the square root is a background normalisation systematic error, arising
from the uncertainty in the determination of $\kappa$.
For a sufficiently large number of events, $\delta
\kappa \sim \kappa/\sqrt B$ is smaller than the assumed
20\% systematic error in the total
cross section. On the other hand, this approach has the drawback that the
significance is determined by $S'$, which may be significantly smaller than $S$
if off-peak signal contributions (combinatorial background) are large, and the
``effective'' statistical error in the background is $\sqrt{\kappa B}$.
Besides, this background rescaling assumes that the main
sources of systematic error ({\em e.g.} $b$ and light jet tagging efficiencies,
jet energy resolution, etc.) do not significantly affect the shape of the
relevant distribution in which the peak is observed.

\begin{figure}[htb]
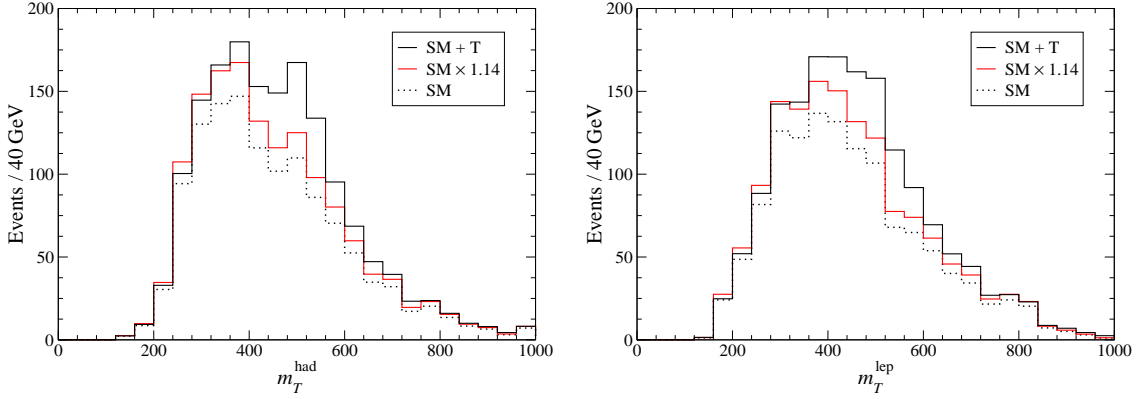

\begin{center}
\begin{tabular}{cc}
\epsfig{file=Figs/mThad-anIII4b-final.eps,height=5.2cm,clip=} &
\epsfig{file=Figs/mTlep-anIII4b-final.eps,height=5.2cm,clip=} 
\end{tabular}
\caption{Reconstructed heavy quark masses
after the kinematical cuts in
Eqs.~(\ref{ec:selIII}).
The dotted lines represent the SM background, and the red lines the same but
rescaled by factor $\kappa \simeq 1.14$.
The continuous lines
correspond to the background plus all heavy quark signals.}
\label{fig:anIIIfinal}
\end{center}
\end{figure}

The probabilistic analysis in section \ref{sec:4.2} is
not the best suited for detecting the peaks in the $\mThad$, $\mTlep$
distributions. Even not including these variables 
in the likelihood functions, requiring a high signal likelihood biases
the background, concentrating the distributions of $\mThad$ and $\mTlep$
around $m_T = 500$ GeV. This is not completely unexpected, since
the signal distributions of the total transverse energy, missing momentum, etc.
have been obtained assuming $m_T = 500$ GeV. Therefore, instead of a
likelihood analysis we perform one based on simple kinematical cuts.
We restrict ourselves to
final states with 4 $b$-tagged jets (in the $5b$ and $6b$ channels the
background rescaling has a larger uncertainty due to the smaller
statistics). We require
\begin{eqnarray}
H_T & \geq & 1000 ~\text{GeV} \,, \notag \\
\ptbmax & \geq & 100 ~\text{GeV} \,, \notag \\
\mathcal{N}_\text{jet} & \leq & 7
\label{ec:selIII}
\end{eqnarray}
to reduce the background. The reconstructed mass distributions obtained are
presented in Fig.~\ref{fig:anIIIfinal}.
The SM background is normalised with
cross section measurements in the regions $160~\text{GeV} \leq \mThad,\mTlep
\leq 360$ GeV, $680~\text{GeV} \leq \mThad,\mTlep
\leq 840$ GeV, obtaining similar rescaling factors in both
distributions, $\kappa=1.139 \pm 0.051$ and $\kappa = 1.141 \pm 0.050$
respectively. Within the mass windows
\begin{equation}
360~\text{GeV} \leq \mThad,\mTlep \leq 640~\text{GeV}
\label{ec:peakIII}
\end{equation}
the significance of the signal (over the rescaled background)
is  $\mathcal{S}_\kappa = 4.27\sigma$. 
(The total number of events after the cuts in Eqs.~(\ref{ec:selIII}) and the
events in the peak regions can be read in Table~\ref{tab:anIII}.)
In this example we find a smaller sensitivity with this method than with the
probabilistic analysis used in section \ref{sec:4.2}, which was
$\mathcal{S}_{20} = 6.93\sigma$.  Nevertheless, it has the
aesthetical advantage of being able to observe the peaks corresponding to the
new quark with unbiased background.

\begin{table}[htb]
\begin{center}
\begin{tabular}{ccccccc}
& $N_\text{cut}$ & $N_\text{peak}$ 
& \quad \quad &
& $N_\text{cut}$ & $N_\text{peak}$
 \\[1mm]
$\TTWH$                       &  193.9  &  137.6  & &
                     $t \bar t b \bar b$           &  223 & 104  \\
$\TTHH$                       &  80.0   &  43.0   & &
                     $t \bar t c \bar c$           &  43  &  20  \\
$\TTZH$                       &  48.7   &  27.2   & &
                     $t \bar t b \bar b EW$        &  17.5 &  7.8  \\
$\TTWZ$                       &  23.7   &  15.5  & &
                     $t \bar t c \bar c EW$        &  1.3   &  1.0 \\
$\TTZZ$                       &  4.5    &  2.2  & &
                     $Wjjjjjj$                     &  4  &  1 \\
$T \bar T b \bar b$           &  1.2    &  0.6  & &
                     $W b \bar b jjjj$             &  33 & 11 \\
$t \bar t H$                  &  26.2   &  10.0  & &
                     $W c \bar c jjjj$             &  6     &  1  \\
$t \bar t$                    &  36     &  15   & &
                     $W b \bar b b \bar b$         &  1  &  1  \\
$t \bar tj $                  &  75     &  33   & &
                     $Zjjjjjj$                     &  0  &  0  \\
$t \bar t2j$                  &  178    &  61   & &
                     $Z b \bar b jjjj$             &  7     &  4  \\
$t \bar t3j$                  &  243    &  105    & &
                     $Z c \bar c jjjj$             &  0  &  0  \\
$t \bar t4j$                  &  222    &  100   & &
                      \\
$t \bar t5j$                  &  135    &  58 
\end{tabular}
\caption{
Number of events with 4 $b$ tags (for 30 fb$^{-1}$) after the selection cuts in
Eqs.~(\ref{ec:selIII}) ($N_\text{cut}$) and also within
the mass windows in Eq.~(\ref{ec:peakIII}) ($N_\text{peak}$).}
\label{tab:anIII}
\end{center}
\end{table}

\section{Other results}

We conclude this analysis examining the dependence of our results on some of our
assumptions. We can estimate how our results change if:  (i) we use MRST
structure functions \cite{mrst}; (ii) we include the charged lepton
identification efficiency; (iii) we select $b$ tagging efficiencies of
50\% or 70\%; (iv) a systematic uncertainty of 30\% is assumed in the
background. In the first case we compute the significances
rescaling the numbers of events in Tables~\ref{tab:anI}--\ref{tab:anII5b6b} by
factors reflecting the change
in the cross sections. In the second case we naively use an
average charged lepton identification efficiency of 90\%.
For the third, we provide crude estimates based on
rescaling by the nominal $b$ tagging efficiencies and rejection factors. 
The resulting significances for the Higgs signals are collected in
Table~\ref{tab:change}. For the $T$ discovery in the $4b$, $5b$ and $6b$
channels they are slightly larger, as shown
in the previous section.

\begin{table}[h*]
\begin{center}
\begin{tabular}{lcccc}
& $t \bar t H$ & $T \bar T (H,4b)$ & $T \bar T (H,5b)$ &
 $T \bar T (H,6b)$ \\
Standard         & 0.39 &  6.43 &  6.02 &  5.63 \\
MRST             & 0.38 &  7.30 &  6.73 &  6.45 \\
$\ell$ eff. 90\% & 0.38 &  6.24 &  5.86 &  5.41 \\
$b$ eff. 50\%    & 0.68 &  6.28 &  5.41 &  4.80 \\
$b$ eff. 70\%    & 0.12 &  1.74 &  1.70 &  1.81 \\
sys 30\%         & 0.31 &  5.08 &  4.72 &  4.78
\end{tabular}
\caption{Estimates of the Higgs signal significances under different
assumptions, explained in the text.}
\label{tab:change}
\end{center}
\end{table}

The results are rather stable except for a 70\% $b$ tagging efficiency, where
backgrounds grow due to the larger mistagging rate.
For a slightly different Higgs mass the results are stable too, as long as the
decay $H \to b \bar b$ dominates, and an additional (small) dependence on $M_H$
is through
the branching ratios for $T$ decays, plotted in Fig.~\ref{fig:mass-BR}.
For larger $T$ masses the signal is suppressed (and for lighter $T$ enhanced)
as a consequence of the variation in the $T \bar T$ cross section,
plotted in Fig.~\ref{fig:mass-cross}. For instance, for a heavy quark mass
$m_T = 600$ GeV the cross section is
776 fb, almost three times smaller than for $m_T = 500$ GeV. On the other hand,
the SM background decreases for larger transverse momenta, but the latter effect
does not make up for the reduction in the $T \bar T$ cross section.

\section{Summary}
\label{sec:6}

Heavy singlet decays were recognised early as an important source
of Higgs bosons \cite{paco}, with a branching ratio close to 25\%
for $M_H \ll m_T$. In this
work we have addressed their experimental observation at LHC, assuming a Higgs
mass of 115 GeV and the possible existence of a 500 GeV heavy quark $T$. We have
performed a detailed signal and background study, with matrix-element-based
generators for the hard processes, subsequent parton showering and hadronisation
by {\tt PYTHIA} and a fast simulation of the ATLAS detector.
As a by-product, new leading-order event generators for
$t \bar t b \bar b$, $t \bar t c \bar c$, $t \bar t H$, $W b \bar b b \bar b$
and other processes have been developed.
These generators include top quark, $W$ and Higgs boson decays and take finite
width and spin effects into account. Their output provides the
colour information necessary for hadronisation.

In our analysis
we have first reevaluated the discovery potential of $t \bar t H$ production,
with $H \to b \bar b$ and semileptonic decay of the $t \bar t$ pair, in the SM.
Our result, $0.4\sigma$ significance for 30 fb$^{-1}$ in low luminosity
running, is similar to the most recent one by CMS,
although the details of the analysis (full simulation for the CMS analysis, with
inclusion of a $K$ factor for $t \bar t H$) differ. Both results are
substantially more
pessimistic than earlier ones \cite{tdr,ttHupd1,ttHupd2}, because in previous
studies only the lowest orders of the leading $\ttnj$ background were taken into
account, and systematic uncertainties in the background normalisation
were not considered.
The $b$ tagging performance has a large impact on the final result, especially
regarding the dominant $\ttnj$ background. We have used the efficiencies
implemented in {\tt ATLFASTB} for the low luminosity run: 60\% $b$ tagging rate
and nominal rejection factors of $6.7$ for charm and $93$ for light jets (with
$p_t$-dependent corrections).
If the latter are better than expected, the observability of $t \bar t H$
production will improve. In this respect, full simulations of
matrix-element-generated signals and backgrounds would be welcome, but it is not
likely that results will attain observability of $t \bar t H$.
Results also depend to some extent on the ability to reconstruct invariant
masses. With a full simulation the mass reconstruction may be degraded,
although studies performed for top pair production have shown good
agreement between fast and full simulations, not only for reconstructed masses
but also for angular distributions \cite{marsella}. On the other hand, it must
be pointed out that our results are conservative in the sense that higher
multiplicity backgrounds $\ttnj$ are included but not higher multiplicity signal
processes $\ttHnj$. The latter might improve the observability by a factor of
two.

New Higgs signals from $T \bar T$ decays,
$T \bar T \to W^+ b \, H \bar t / H t \, W^- \bar b$,
$T \bar T \to H t \, H \bar t$ and
$T \bar T \to Z t \, H \bar t / H t \, Z \bar t$,
have been then examined.
We have demonstrated
that, in a standard search for $t \bar t H$ production, a possible contribution
of these processes can easily be overlooked, and do not much improve the Higgs
observability. We have presented a novel reconstruction technique specific to
the search for the leading signals
$T \bar T \to W^+ b \, H \bar t / H t \, W^- \bar b \to W^+ b W^- \bar b H$,
$T \bar T \to H t \, H \bar t \to W^+ b W^- \bar b H H$,
which does not require knowledge of the heavy quark
mass. Despite their different kinematics and large
transverse momenta, these signals are not easy to isolate from the $\ttnj$
background, which is large and also involves larger transverse momenta
for increasing values
of $n$. Using a likelihood analysis, these processes are cleanly separated from
the SM background, giving a high statistical significance for the Higgs,
$10.4\sigma$ for 30 fb$^{-1}$ including a 20\% systematic uncertainty in
the background normalisation. In the case that a 500 GeV $T$ quark
exists, 8 fb$^{-1}$ of luminosity could suffice to discover the Higgs
boson. This striking signal
is due to the large $T \bar T$ production cross section (2.14 pb for $m_T = 500$
GeV), the large branching ratio for final states with Higgs bosons,
$\text{Br}(T \bar T \to H + X) = 0.55$, and the distinctive features
of these processes: in addition to larger transverse momenta, a high $b$ jet
multiplicity in the final state and reconstructed invariant masses peaking at
$m_T$.

Finally, we have  addressed the observability of the new quark, which is not
equivalent to the discovery of the Higgs boson because the classification of
processes as signals and background differs. We have shown that a significance
of $11.7 \sigma$ is reached for 30 fb$^{-1}$, similar to the one in the
$T \bar T \to W^+ b W^- \bar b$ channel (a detailed comparison between both
channels is difficult because of the different assumptions made in the two
studies). We have also used a standard analysis
in order to show that the peaks in the invariant mass distributions of the
heavy quarks would be easy
to observe, even considering the uncertainties in the background normalisation.
For higher $T$
masses, $T \bar T \to W^+ b W^- \bar b$ is the leading discovery channel,
due to three facts: (i) the branching ratio for $\ell \nu bbbbjj$ final states
decreases slightly with $m_T$; (ii) for heavier $T$, the charged lepton from the
semileptonic decay $T \to W^+ b \to \ell^+ \nu b$ (or the charge conjugate)
generically has a very large transverse momentum which can be exploited to
reduce backgrounds very
efficiently \cite{top06}; (iii) larger $T$ masses can only be explored in a
high luminosity LHC run, where $b$ tagging performance is degraded and multi-jet
backgrounds to the $T \bar T (WH,HH,ZH)$ signals are larger.

\vspace{1cm}
\noindent
{\Large \bf Acknowledgements}
\vspace{0.3cm}

\noindent
I thank F. del Aguila and R. Pittau for useful discussions and for reading the
manuscript, and the referee for many useful suggestions.
This work has been supported by a MEC Ramon y Cajal contract and project
FPA 2003-09298-C02-01, and by
Junta de Andaluc{\'\i}a projects FQM 101 and FQM 437.

\appendix

\section{Probabilistic analysis}
\label{ap:b}

In the probabilistic analysis we build likelihood
functions which use information from several kinematical variables
to discriminate between event classes, namely the signal (one or more) and the
background.
For a given kinematical variable $x$, {\em e.g.} a transverse momentum,
different event classes $j=1,\dots,m$ have different kinematical
distributions $f^j (x)$, which we normalise to unity. We define the
``probability'' function
\begin{equation}
p_j(x) = \frac{f_j(x)}{\sum_k f_k(x)} \,.
\end{equation}
If the distributions $f^j$ are normalised to their total cross section, the
function $p^j(x)$
represents the probability that the event corresponds to the class $j$, and
when normalised to unity $p^j$ it is the relative probability (up to total
cross section factors). 
For a set of kinematical variables $x_i$, $i=1,\dots,n$,
the likelihoods $L_j$ are then defined as the product of the probabilities for
each variable $x_i$,
\begin{equation}
L_j(x_1,\dots,x_{n}) = \prod_{i=1}^n p_j^i(x_i) \quad j=1,\dots,m \,.
\end{equation}
Selection cuts may be applied on likelihood ratios $L_{S_i}/L_B$, for $S_i$
and $B$ the signal and background classes, respectively, in order to enhance the
signal(s). Alternatively,
instead of working directly with these ratios it is often more
practical to consider the logarithm of these quantities,
$\log_{10} L_{S_i}/L_B$.

\begin{figure}[htb]
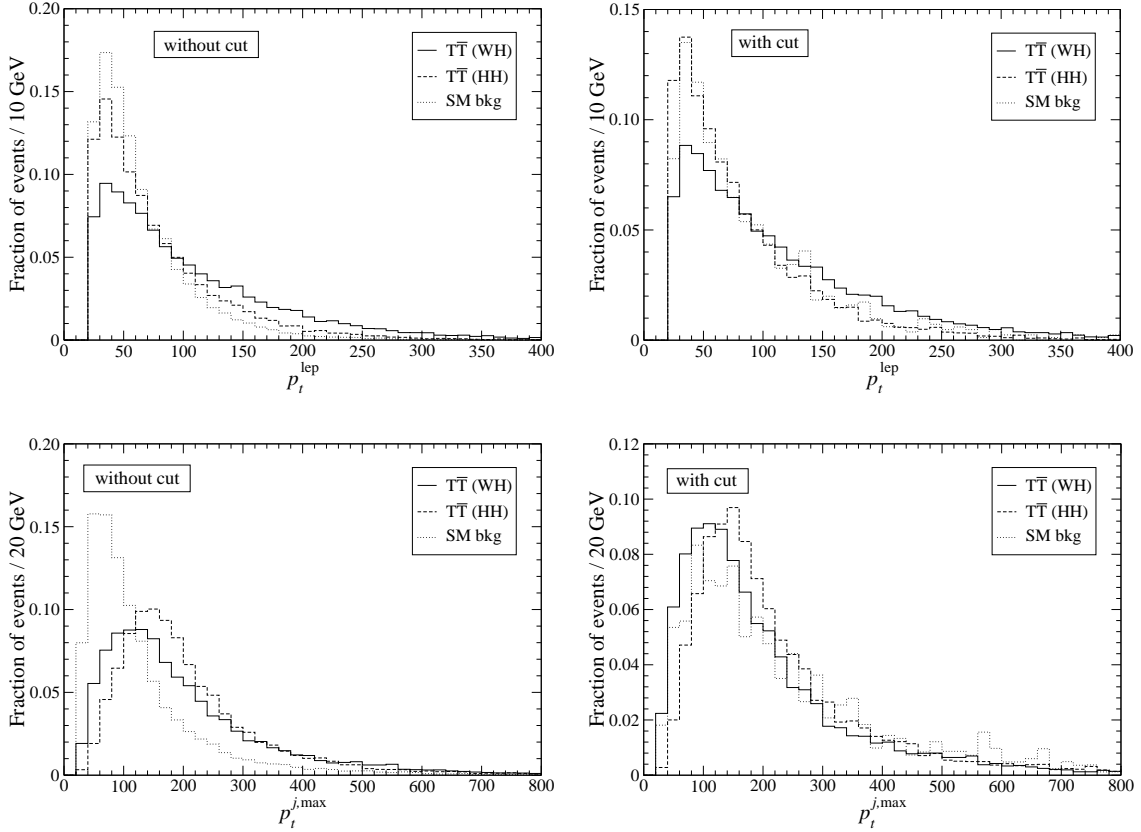

\begin{center}
\begin{tabular}{cc}
\epsfig{file=Figs/ptlep_n-ap.eps,height=5.2cm,clip=} &
\epsfig{file=Figs/ptlepc_n-ap.eps,height=5.2cm,clip=} \\[0.4cm]
\epsfig{file=Figs/ptlmax_n-ap.eps,height=5.2cm,clip=} &
\epsfig{file=Figs/ptlmaxc_n-ap.eps,height=5.2cm,clip=} 
\end{tabular}
\caption{Normalised variables $\ptlep$, $\ptlmax$ for the analysis II ($4b$
final states), without cuts and after requiring $\log_{10} L_S/L_B \geq 2$, with
$L_S$, $L_B$ involving the rest of variables.}
\label{fig:ap}
\end{center}
\end{figure}

We emphasise that performing a probabilistic analysis of this type is not as
straightforward as one might think. Naively, one would take all the relevant
variables which exhibit different distributions for signal and background and
build with them likelihood functions. But this is not optimal and, perhaps
surprisingly, some variables which one might consider as relevant actually
reduce the discriminating power of the likelihood functions. This can be
understood as a result of the fact that some variables are correlated, and selecting values of one of them
modifies the distribution of the others. Let us take as example the
transverse momentum distributions of the charged lepton ($\ptlep$) and the light
jet with maximum $p_t$ ($\ptlmax$) for the analysis II in $4b$ final states.
These variables have not been included in the probabilistic analysis in this
case. Their
normalised distributions before any cut are presented in Fig.~\ref{fig:ap}
(left), and after requiring $\log_{10} L_S/L_B > 2$ (but without them on the
likelihood functions) on the right. From their distributions in the
left column we observe that their inclusion in the likelihood functions
would favour larger transverse momenta,
since the background distributions are peaked at lower $p_t$. But observing the
right column we realise that this would  actually disfavour the signal over the
background (for example, the tail in the $\ptmax$ distribution {\em after the
likelihood cut} is larger for the background than for the two signals, and in
the $\ptlep$ distribution larger for the background than for $\TTHH$).
These
examples make apparent that optimising the analysis requires educated guessing
and trial and error to find (or get close to) the best set of variables.


\begin{thebibliography}{99}

\bibitem{higgs}
P.~W.~Higgs, {\em Broken Symmetries And The Masses Of Gauge Bosons,
Phys. Rev. Lett.}  {\bf 13} (1964) 508;
{\em Spontaneous Symmetry Breakdown Without Massless Bosons,
Phys. Rev.}  {\bf 145} (1966) 1156.

\bibitem{hlimit}
R. Barate {\it et al.}  [LEP Working Group for Higgs boson searches],
{\em Search for the standard model Higgs boson at LEP,
Phys. Lett.} {\bf B 565} (2003) 61
[{\tt hep-ex/0306033}].

\bibitem{hfit}
LEP Electroweak Working group,
{\em A combination of preliminary electroweak measurements and constraints on
the standard model,}
{\tt hep-ex/0511027};
see also \\ {\tt http://lepewwg.web.cern.ch/LEPEWWG}

\bibitem{hlimit2}
N. Cabibbo, L. Maiani, G. Parisi and R. Petronzio,
{\em Bounds On The Fermions And Higgs Boson Masses In Grand Unified Theories,
Nucl. Phys.} {\bf B 158} (1979) 295.

\bibitem{tdr}
ATLAS collaboration, 
{\em ATLAS detector and physics performance technical design report},
CERN-LHCC-99-15.

\bibitem{CMS}
CMS collaboration, 
{\em CMS Physics technical design report volume II: Physics Performance},
CERN-LHCC-2006-021.

\bibitem{ttHupd1}
J. Cammin and M. Schumacher,
{\em The ATLAS discovery potential for the channel $t \bar t H$, $H \to b \bar
b$},
ATLAS note ATL-PHYS-2003-024.

\bibitem{ttHupd2}
B. King, S. Maxfield and J. Vossebeld,
{\em Search for a light Standard Model Higgs in the channels $pp \to t \bar t
H,WH,ZH$},
ATLAS note ATL-PHYS-2004-031.

\bibitem{cmshiggs}
S. Cucciarelli {\em et al.},
{em Search for $H \to b \bar b$ in association with a $t \bar t$ pair at CMS},
CMS note 2006/119

\bibitem{VBF}
S.~Asai {\it et al.},
{\em Prospects for the search for a standard model Higgs boson in ATLAS  using
vector boson fusion,
Eur. Phys. J.} {\bf C 32S2} (2004) 19
[{\tt hep-ph/0402254}].

\bibitem{WW160}
K. Jakobs and T. Trefzger,
{\em SM Higgs Searches for $H \to WW^{(*)} \to \ell^+ \nu \ell^- \nu$ with a
Mass between $150-190$ GeV at LHC},
ATLAS note ATL-PHYS-2000-015.

\bibitem{ZZupd}
B. Mohn and B. Stugu,
{\em Corrections to the discovery potential for finding the Standard Model Higgs
in the four lepton final state},
ATLAS note ATL-PHYS-2004-014.

\bibitem{cpyuan}
C.~R.~Chen, K.~Tobe and C.~P.~Yuan,
{\em Higgs boson production and decay in little Higgs models with T-parity,}
{\tt hep-ph/0602211}.

\bibitem{paco}
F. del Aguila, G. L. Kane and M. Quiros,
{\em A Possible Method To Produce And Detect Higgs Bosons At Hadron Colliders,
Phys. Rev. Lett.} {\bf 63} (1989) 942;
F. del Aguila, L. Ametller, G. L. Kane and J. Vidal,
{\em Vector Like Fermion And Standard Higgs Production At Hadron Colliders,
Nucl. Phys.} {\bf B 334} (1990) 1.

\bibitem{lhiggs}
N. Arkani-Hamed, A. G.~Cohen and H. Georgi,
{\em Electroweak symmetry breaking from dimensional deconstruction,
Phys. Lett.} {\bf B 513} (2001) 232 [{\tt hep-ph/0105239}];
N. Arkani-Hamed, A. G. Cohen, E. Katz and A. E. Nelson,
{\em The littlest Higgs,
JHEP} {\bf 0207} (2002) 034 [{\tt hep-ph/0206021}].

\bibitem{extrad}
E. A. Mirabelli and M. Schmaltz,
{\em Yukawa hierarchies from split fermions in extra dimensions,
Phys. Rev.} {\bf D 61} (2000) 113011 [{\tt hep-ph/9912265}];
S. Chang, J. Hisano, H. Nakano, N. Okada and M. Yamaguchi,
{\em Bulk standard model in the Randall-Sundrum background,
Phys. Rev.} {\bf D 62} (2000) 084025 [{\tt hep-ph/9912498}];
F. del Aguila and J. Santiago,
{\em Universality limits on bulk fermions,
Phys. Lett.} {\bf B 493} (2000) 175 [{\tt hep-ph/0008143}].


\bibitem{frampton}
P. H. Frampton, P. Q. Hung and M. Sher,
{\em Quarks and leptons beyond the third generation,
Phys. Rept.} {\bf 330} (2000) 263 [{\tt hep-ph/9903387}].

\bibitem{plb}
J. A. Aguilar-Saavedra,
{\em Pair production of heavy $Q = 2/3$ singlets at LHC,
Phys. Lett.} {\bf B 625} (2005) 234
[{\em Erratum-ibid.} {\bf B 633} (2006) 792] [{\tt hep-ph/0506187}].

\bibitem{unel}
R. Mehdiyev, S. Sultansoy, G. Unel and M. Yilmaz,
{\em Search for E6 isosinglet quarks in ATLAS,}
{\tt hep-ex/0603005.}

\bibitem{4gen}
E. Arik, M. Arik, S. A. Cetin, T. Conka, A. Mailov and S. Sultansoy,
{\em With four standard families, the LHC could discover the Higgs boson with a
few fb$^{-1}$, Eur. Phys. J.} {\bf C 26} (2002) 9 [{\tt hep-ph/0109037}].

\bibitem{paconpb}
F. del Aguila and M. J. Bowick,
{\em The Possibility Of New Fermions With $\Delta I = 0$ Mass,
Nucl. Phys.} {\bf B 224} (1983) 107.

\bibitem{london}
P. Langacker and D. London,
{\em Mixing Between Ordinary And Exotic Fermions,
Phys. Rev.} {\bf D 38} (1988) 886.

\bibitem{barger}
V. D. Barger, M. S. Berger and R. J. N. Phillips,
{\em Quark singlets: Implications and constraints,
Phys. Rev.} {\bf D 52} (1995) 1663 [{\tt hep-ph/9503204}].

\bibitem{largo}
J. A. Aguilar-Saavedra,
{\em Effects of mixing with quark singlets,
 Phys. Rev.} {\bf D 67} (2003) 035003
[{\em Erratum-ibid.} {\bf D 69} (2004) 099901] [{\tt hep-ph/0210112}].

\bibitem{cdf}
CDF Collaboration,
{\em Search for heavy Top $t' \to W q$ in Lepton Plus Jets events},
CDF note 8003; see also \\
{\tt http://www-cdf.fnal.gov/physics/new/top/2005/ljets/tprime/gen6/public.html}

\bibitem{pdb}
W.~M.~Yao {\it et al.}  [Particle Data Group],
{\em Review of particle physics,
J. Phys.} {\bf G 33} (2006) 1.

\bibitem{prl}
F. del Aguila, J. A. Aguilar-Saavedra and R. Miquel,
{\em Constraints on top couplings in models with exotic quarks,
Phys. Rev. Lett.} {\bf 82} (1999) 1628 [{\tt hep-ph/9808400}].

\bibitem{azuelos}
G. Azuelos {\it et al.},
{\em Exploring little Higgs models with ATLAS at the LHC,
Eur. Phys. J.} {\bf C 39S2} (2005) 13  [{\tt hep-ph/0402037}].

\bibitem{pythia}
T. Sjostrand, S. Mrenna and P. Skands,
{\em PYTHIA 6.4 physics and manual},
JHEP {\bf 0605} (2006) 026
[{\tt hep-ph/0603175}].

\bibitem{alpgen}
M. L. Mangano, M. Moretti, F. Piccinini, R. Pittau and A. D. Polosa,
{\em ALPGEN, a generator for hard multiparton processes in hadronic collisions,
JHEP} {\bf 0307} (2003) 001 [{\tt hep-ph/0206293}];
See also {\tt http://mlm.home.cern.ch/m/mlm/www/alpgen/}

\bibitem{mlm}
M. L. Mangano, talk at Lund University,
{\tt http://cern.ch/~mlm/talks/lund-alpgen.pdf}

\bibitem{tauola}
S. Jadach, Z. Was, R. Decker and J. H. Kuhn,
{\em The $\tau$ decay library TAUOLA: Version 2.4,
Comput. Phys. Commun.} {\bf 76} (1993) 361.

\bibitem{photos}
E. Barberio, B. van Eijk and Z. Was,
{\em Photos: A Universal Monte Carlo For QED Radiative Corrections In Decays,
Comput. Phys. Commun.} {\bf 66} (1991) 115.

\bibitem{atlfast}
E. Richter-Was, D. Froidevaux and L. Poggioli,
{\em ATLFAST 2.0 a fast simulation package for ATLAS},
ATLAS note ATL-PHYS-98-131.

\bibitem{helas}
E. Murayama, I. Watanabe and K. Hagiwara,
{\em HELAS: HELicity Amplitude Subroutines for Feynman Diagram Evaluations},
KEK report 91-11, January 1992.

\bibitem{madgraph}
T. Stelzer and W. F. Long,
{\em Automatic generation of tree level helicity amplitudes,
Comput. Phys. Commun.} {\bf 81} (1994) 357 [{\tt hep-ph/9401258}].

\bibitem{acermc}
B. P. Kersevan and E. Richter-Was,
{\em The Monte Carlo event generator AcerMC version 2.0 with interfaces to
PYTHIA 6.2 and HERWIG 6.5,} {\tt hep-ph/0405247}.

\bibitem{vegas}
G. P. Lepage,
{\em Vegas: An Adaptive Multidimensional Integration Program},
Report CLNS-80/447.

\bibitem{cteq}
H. L. Lai {\it et al.}  [CTEQ Collaboration],
{\em Global QCD analysis of parton structure of the nucleon: CTEQ5 parton
distributions,
Eur. Phys. J.} {\bf C 12} (2000) 375 [{\tt hep-ph/9903282}].

\bibitem{epsb}
R. Barate {\em et al.} [ALEPH Collaboration],
{\em Studies of quantum chromodynamics with the ALEPH detector,
Phys. Rept.} {\bf 294} (1998) 1.

\bibitem{int}
A. I. Etienvre, J. P. Meyer and J. Schwindling,
{\em Top quark mass measurement in the lepton plus jet channel using full
simulation},
ATLAS note ATL-PHYS-INT-2005-002.

\bibitem{marsella}
F. Hubaut, E. Monnier, P. Pralavorio, B. Resende and C. Zhu,
{\em Comparison between full and fast simulations in top physics},
ATLAS note ATL-PHYS-PUB-2006-017.

\bibitem{marsella2}
For a recent comparison see
F. Hubaut, E. Monnier, P. Pralavorio, B. Resende and C. Zhu,
{\em Polarization studies in $t \bar t$ semileptonic events with ATLAS full
simulation},
ATLAS note ATL-PHYS-PUB-2006-022.

\bibitem{CMS2}
CMS collaboration, 
{\em CMS Physics technical design report volume I: Detector Performance and
Software},
CERN-LHCC-2006-001.

\bibitem{trigger}
T. Sch\"orner-Sadenius and S. Tapprogge,
{\em ATLAS Trigger Menus for the LHC Start-up Phase},
ATLAS note ATL-DAQ-2003-004.


\bibitem{top06}
J. A. Aguilar-Saavedra,
{\em New signals in pair production of heavy $Q = 2/3$ singlets at LHC},
{\tt hep-ph/0603199}.

\bibitem{baur}
U.~Baur and E.~L.~Berger,
{\em Probing The Weak Boson Sector In Z Gamma Production At Hadron Colliders,
Phys.\ Rev.} {\bf D 47} (1993) 4889.

\bibitem{mrst}
A. D. Martin, R. G. Roberts, W. J. Stirling and R. S. Thorne,
{\em Physical gluons and high-$E_T$ jets,
Phys. Lett.} {\bf B 604} (2004) 61 [{\tt hep-ph/0410230}].







\end{thebibliography}
\end{document}